\documentclass[manuscript]{article} 

\AtBeginDocument{%
  \providecommand\BibTeX{{%
    \normalfont B\kern-0.5em{\scshape i\kern-0.25em b}\kern-0.8em\TeX}}}

\usepackage{array,booktabs}
\usepackage{amsmath,amsfonts,amsthm,amssymb}
\usepackage{bm}
\usepackage{graphicx}
\graphicspath{{./}{Charts/}} %
\usepackage{subcaption}
\usepackage{xspace}
\usepackage{algorithm}
\usepackage{todonotes} 
\usepackage[numbers,sort]{natbib}
\usepackage[noend]{algpseudocode} 
\usepackage{hyperref}
\usepackage{paralist} 
\usepackage{graphbox} 
\usepackage[percent]{overpic} 
\usepackage{transparent} 
\usepackage{balance}
\usepackage{array,multirow}
\usepackage[normalem]{ulem}
\usepackage{wrapfig}
\usepackage{stmaryrd}
\usepackage[margin=1in]{geometry}

\newcommand{\infovec}{information access signature}
\newcommand{\Exp}{\mathbb{E}}
\newtheorem{theorem}{Theorem}

\newtheorem{definition}[theorem]{Definition}

\usepackage[final,inline,nomargin,index]{fixme}

\fxsetup{theme=color,mode=multiuser,inlineface=\itshape,envface=\itshape}
\FXRegisterAuthor{ab}{aab}{\colorbox{green!10!white}{\color{black}Ashkan}}
\FXRegisterAuthor{sv}{asv}{\colorbox{gray!10!white}{\color{black}Suresh}}
\FXRegisterAuthor{sf}{asf}{\colorbox{blue!10!white}{\color{black}Sorelle}}
\FXRegisterAuthor{cs}{acs}{\colorbox{red!10!white}{\color{black}Carlos}}
\FXRegisterAuthor{lk}{alk}{\colorbox{purple!10!white}{\color{black}Nasanbayar}}
\fxusetargetlayout{color}

\begin{document}

\title{Information access representations and social capital in networks\thanks{This research was funded in part by the NSF under grants IIS-1955321 and IIS-1956286.}}

\author{Ashkan Bashardoust\\
  {University of Utah}\\
  {Salt Lake City}, {UT}, {USA}\\
\url{ashkanb@cs.utah.edu} 
\and
Hannah C. Beilinson\\
Haverford College\\
Haverford, PA, USA\\
\url{hcbeilinson@gmail.com}
\and
Sorelle A. Friedler\\
  {Haverford College}\\
  {Haverford},
  {PA}, {USA}\\
\url{sorelle@cs.haverford.edu}
\and 
Jiajie Ma\\
  {Duke University}\\
  {Durham}, {NC}, {USA}\\
\url{jason.ma@duke.edu}
\and
Jade Rousseau\\
  {Haverford College}\\
  {Haverford}, {PA}, {USA}\\
\url{jade.rousseau.paris@gmail.com}
\and
Carlos E. Scheidegger\\
  {Posit PBC}\\
  {USA}\\
\url{carlos.scheidegger@posit.co}
\and
Blair D. Sullivan\\
{University of Utah}\\
{Salt Lake City}, UT, USA\\
\url{sullivan@cs.utah.edu}
\and
Nasanbayar Ulzii-Orshikh\\
University of Michigan\\
Ann Arbor, MI, USA\\
\url{nulziior@umich.edu}
\and
Suresh Venkatasubramanian\\
  {Brown University}\\
  {Providence},
  {RI},
  {USA}\\
\url{suresh@brown.edu}
}



\maketitle

\begin{abstract}
Social network position confers power and social capital. In the setting of online social networks that have massive reach, creating mathematical representations of social capital is an important step towards understanding how network position can differentially confer advantage to different groups and how network position can itself be a source of advantage. In this paper, we use well established models for information flow on networks as a base to propose a formal descriptor of the network position of a node as represented by its information access. Combining these descriptors allows a full representation of social capital across the network. Using real-world networks, we demonstrate that this representation allows the identification of differences between groups based on network specific measures of inequality of access.
\end{abstract}

\section{Introduction}
\label{sec:introduction}

It has been known for decades that belonging and positionality in a social network confer power to individuals in the form of \emph{social capital} \citep{hethcote2000mathematics, burt, burt2004holes, burt2000socialcapital, coleman, coleman1988social, granovetter78, jackson19human, podolny97resources}. More recently, information access has appeared as one of, if not \emph{the}, key component of social capital. On the one hand, information access is structural in nature such that what information one has access to is determined by who one knows, by one's place in social networks. On the other hand, what information one has access to (in a broad sense of the term) is so determining for navigating the social world that position in a network is itself a first-order ``feature'' that can lead to discrimination in and of itself, separately from (while often compounded with) individual demographic features \cite{boyd14Nature}. Both the structural nature of information access and its relation to overall individual social capital is exacerbated in \emph{online} social networks, where curation processes -- recommendations most notably -- encourage the formation of self-preferential links between people (nodes) and thereby reinforce existing information advantage and help create classes of people based on their relative access to information. Node groups based on information access might therefore be more salient to understanding network privilege dynamics than traditional groups based on node-level features like demographics (a.k.a sensitive attributes). 
It is this observation that motivates this work. Our goal is to formalize a representation of social capital, based on information access, for a network such that individuals' social position and associated group structure can be studied.

\subsection{Contributions}
\label{sec:contributions}
In this paper, we:
\begin{itemize}
\item connect the theory of social capital and information access to an introduced mathematical representation;
\item formalize the notion of an \emph{information access representation} of a network;
\item show how a full representation can be effectively computed for a variety of network sizes;
\item and validate the representation by showing that clustering nodes based on information access captures external measures of real-world interest through experiments on real-world networks.
\end{itemize}
We also present a formal interpretation of information access representations and distinguish it theoretically and empirically from existing methods of identifying structure in networks, illustrating the novelty and value of the information access lens. 


\section{Related Work}
\label{sec:related-work}

Inspired by the literature on social position initiated by \citep{granovetter1973} and framed in the context of online social networks by \citep{boyd14Nature}, there has been a recent development of computational questions around fairness in access on social networks \cite{fish2019gaps, tsang2019group, rahmattalabi, ali2019fairness, jalali2020unfairness, Becker_DAngelo_Ghobadi_Gilbert_2021, becker2023improving}. The starting point for all of this is the idea of \emph{information access as a resource}. In \citep{fish2019gaps}, this is formalized in terms of the question ``what is the probability that vertex $i$ in a graph gets information from source $j$,'' and the paper focuses on explicit interventions (in the form of augmenting the seed set) to ensure that the minimum such probability is maximized.
Subsequent research \cite{tsang2019group, stoica2019fairness} has focused on the problem of allocation with respect to information access -- attempting to make sure that different demographic groups (represented as disjoint subsets of the graph) receive similar information access. Recently, this work has expanded to include structural notions of such fairness of access \cite{jalali2022fairness, mehrotra2022revisiting}, still with a focus on demographic groups.
Information access itself draws on the mechanisms of influence maximization, which was studied initially by \citep{domingos01mining} and formalized by \citep{kempe03maximizing}, leading to an extensive literature on the subject \citep{li2018influence}. Related literature that is not directly connected to this work includes an examination of network formation biases (through natural mechanisms and through
automated recommendations in online social networks) might exacerbate inequity
(in the form of reduced social capital) and reduced opportunities for those in
the out-groups~\cite{stoica2019fairness, DBLP:journals/pomacs/ZhangHMBC21, oliveira2021mixing, lee2019homophily, karimi2018homophily, morgan2018prestige, clauset:etal:2015, wang2021information}.
Another line of work~\cite{bashardoust2023reducing, jalali2020unfairness, FairEdge22Swift} consider adding edges instead of seeds to improve fairness in networks. In particular, Bashardoust et al. measures the structural advantage of a node based on the access signatures~\cite{bashardoust2023reducing}, referencing the originally preprinted version of this paper~\cite{beilinson2020clustering}.

The idea of using social processes of group formation in a method for finding \emph{clusters} in a graph, while distinct from our information-access-based framework, is generally referred to as \emph{community discovery}. A slightly different perspective often referred to as \emph{role-based} discovery also seeks to recognize that groups may not be identified based on proximity, but based on similar \emph{roles} that entities play in the network. For example, Henderson et al. propose role extraction models in large graphs~\cite{Henderson11Who, Henderson12Rolx}. These are not mutually exclusive notions: in practice one way to ``interpolate'' between the two approaches is to decide the size of neighborhood that is relevant when determining if two nodes have similar roles or are in the same community. See \citep{rossi20tkdd-roles} for a recent survey of this literature. Another line of work recognizes that graphs manifest a variety of structures at the \emph{mesoscale}\citep{fortunato2010community} -- when we are not merely looking at either individual node neighborhoods or aggregate properties of the entire graph. Thus, a fruitful strategy is to focus on specific structures of interest and detect them directly. One such pattern that is particularly relevant for this paper is \emph{core-periphery} structures \citep{borgatti2000models,rombach2014core,csermely2013structure} described by a central well-connected core that links with a number of disconnected pieces.  We explore connections to these methods further in Section \ref{sec:comparisons}.

Moving further afield, graph clustering itself has long been a focus of intense study (see \cite{Aggarwal2010} for a survey). While a detailed review of the different strands of graph clustering is beyond the scope of this paper, one can categorize graph clustering algorithms as those based on finding dense submotifs in a graph, those based on spectral analysis \cite{shi2000normalized,von2007tutorial,gharan2014partitioning,10.1145/990308.990313} (which in turn generalizes connectivity-based clusterings) and those based on the more general framework of unusual local density or modularity \cite{brandes2007modularity}.

Recent research has delved into the examination of the information access challenge within the realm of theoretical graph neural networks (GNNs)~\cite{Alon2021, Topping2022, Karhadkar2023, Arnaiz2022, DiGiovanni2023}. Within this body of work, a number of studies propose the utilization of random-walk spectral metrics as a means to effectively characterize the dynamics of information propagation~\cite{Topping2022, Arnaiz2022, Velingker2022, DiGiovanni2023}. Additionally, Dong et al. address the problem of explaining and mitigating the structural bias of input networks in GNNs~\cite{Dong2022Structural, Dong2022Modeling}.


\section{Information Access: Social Theory}
\label{sec:social-motivation}
	\textit{Social capital} 
	refers to the value that is gained from ‘being social’; that is, the idea that 
	being a part of social structures is a source of benefit, utility, power; and, further, one's place within these social structures is itself a determinant of power in part because it determines access to various resources \cite{kaldis2013encyclopedia}. 
	Most of us likely already have an intuitive idea of the significance of belonging and positionality when it comes to access to information---e.g., who you are friends with determines whether you will be aware that an event will occur. This access to information is a resource determined by our 
	relations with others \cite{huysman2004social}. 
	
	Position in a network is thus a first-order `feature'. As such, it can in and of itself lead to discrimination though in reality, it is intertwined with other first order features such as individual demographics \cite{boyd14Nature}. Such an emphasis on position in a network is characteristic of the \emph{network approach} to social capital, which developed following Granovetter’s 1973 seminal paper \cite{granovetter1973}. This approach tends to focus on \emph{structural social capital}: looking at the properties of the social system and of the network of relations as a whole, and approaching social capital based on what relationships (edges) reveal about a chosen proxy of social capital \cite{claridge2020structural, claridge2020measure}.
	We take this network approach to social capital, and acknowledge in doing so that we do not examine aspects of social capital that may not be visible through a focus on and examination of network structure.
	Information is also 
	a locus of power, and various studies have 
	linked deep-rooted inequalities to disparate levels of access to information \cite{mikiewicz2021social}, such as racial income differences in the US \cite{huysman2004social}. 
	Information exists and is accessed \textit{within} networks and one’s relationship to information is thus a crucial element of one’s social capital; a relationship which network analysis is uniquely positioned to investigate. 
 
    Furthermore, information is not merely spread through a social structure, it itself participates in the creation of that structure \cite{balnaves1993sociology}. 
    In online social networks, 
    not only is information (in the form of content) essentially what these platforms are about, 
    but the users themselves are, in an important way, defined in terms of their relationship to information: which content a user is exposed to, which information they consume, how they consume it, and so on itself becomes information as \emph{data} \cite{mosseri2021instagram}. We thus argue that information access can be conceptualized and analyzed not only as a resource an individual has or hasn't access to, but as a defining feature of an individual. 
    Our claim 
    is consistent with  
    a cultural structuralist conceptualization of social capital, which views social capital as not merely something an individual \textit{has}, but as \textit{a part of} an individual~\cite{siisiainen2003two, claridge2020cultural}. 
 
	We therefore propose to represent the social capital of a node $i$ in a network in terms of its \emph{information access signature}: a vector containing the probabilities of node $i$ accessing information from each other node in the network $j$ which spreads it. Introduced formally in Section \ref{sec:representation}, information access signatures for all nodes can be combined into a matrix representation of the information access of the full network. Because the matrix is symmetrical along the diagonal in an undirected network, the vector may also be thought of in terms of `influence' (or `contagion' , `diffusion'): that is, as containing the probabilities of node $i$ spreading an information to each other node $j$. 	The information access signature thus \textit{structurally} defines a node in terms of the map of its access to information. 
	For two nodes to have a similar information signature thus means that they have a similar pattern of connection, which in turn means that they are \textit{structurally equivalent}. The concept of structural equivalence was developed in sociology to explain the mechanisms of diffusion, with the goals of understanding how information comes to be spread or how a practice comes to be adopted, and is a characteristic of nodes that occupy similar positions and roles in a network \cite{adam2008using}. Structurally equivalent nodes do not need to have linkages to each other, nor do they need to be connected to the same nodes for them to share a similar structural position; what matters is the \textit{pattern} of connections \cite{liu2017social}. For example, one might draw strong parallels between the conditions of people living in different cities, though they inhabit different networks, because of similarities in their \textit{structural embedding}. We thus claim that the information access representation of the network is a mathematical representation of social capital that allows examination of structurally equivalent individuals or groups in a network, e.g. via clustering. We 
	contend that since information access is both a critical first-order feature and a feature within which other features, such as demographics, are embedded, this representation offers a novel and valuable way of looking at communities and inequalities, problems and solutions.


\section{Information Access: Mathematical Representation, Clustering, and Analyses}
\label{sec:definitions}

Let $G = (V, E)$ be a network with sets of nodes $V$ and edges $E$, where $|V| =
n$. Consider any  \emph{information flow model} that describes how information
might transmit from one node to its neighbors, such as the independent cascade
model, the linear threshold model, or any of the infection flow models from
epidemiology. All these models are stochastic and assume some initial
\emph{seed} set of nodes that possess the information to be spread. For any given seed set $S$  
there is then a fixed probability $p_{v, S}$ that node $v \in V$ possesses the 
information once the process terminates. 
This motivates our idea of an \infovec: a way to encode the
``view'' from a node $v$ of the access it has to information sent from other
nodes in the graph.

\subsection{Information access signatures, representation, and clustering}
\label{sec:representation}

\begin{definition}[\infovec]
The \infovec\ $s^G_{\bm{\alpha}} \colon V \to \mathbb{R}^n$ for $v_j \in V$ is 
$ s^G_{\bm{\alpha}}(v_j) = (p_{1j}, \dots, p_{ij}, \dots p_{nj}) $
where $p_{ij}$ is the probability that node $v_j \in V$ receives information
seeded at node $v_i \in V$ under some information flow model with parameter vector $\bm{\alpha}$. 
\end{definition}

 Intuitively, this signature characterizes a node's information access based on how likely they are to receive information from everyone else in the network; people who are likely to receive information from the same part of the network will have similar signatures.  Given our understanding of information access as a measure of network privilege, this means two nodes with similar access signatures should have comparable information privilege. 
 
 We can use these signatures to define an information access representation of the network.

\begin{definition}[information access representation]
The information access representation for $v_j \in V$ is
\[ R^G_{\bm{\alpha}} = \{~ s^G_{\bm{\alpha}}(v_j) ~|~ v_j \in V ~\} \]
where $\bm{\alpha}$ represents the parameters of the information flow model as
before. 
\end{definition}

The information access representation $R^G_{\bm{\alpha}}$ represents each vertex
of the graph as a point in an $n$-dimensional space. Two points are close in
this space if they have similar views of information flow to/from other nodes, i.e. similar information access privilege or social capital. This thus allows us to group together nodes based on this proximity, i.e. to compute a graph clustering.

Formally, we will endow this space with an $\ell_2$ norm, which then allows us
to use standard clustering techniques. Throughout this work, we'll use
$k$-means, though other standard clustering techniques can be easily applied to
the representation.  We will call the clustering resulting from applying
$k$-means clustering to the information access representation for a given
$\alpha$ and $k$ desired clusters the \emph{information access clustering} of a
network.

\begin{figure}[t]
  \centering
  \includegraphics[width=1.5in]{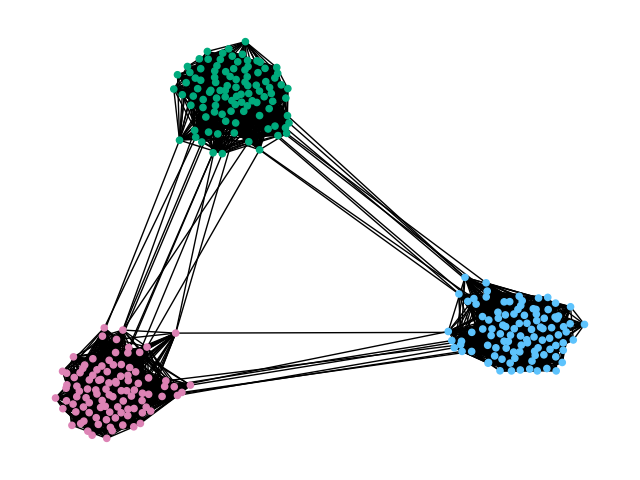}
  \includegraphics[width=1.5in]{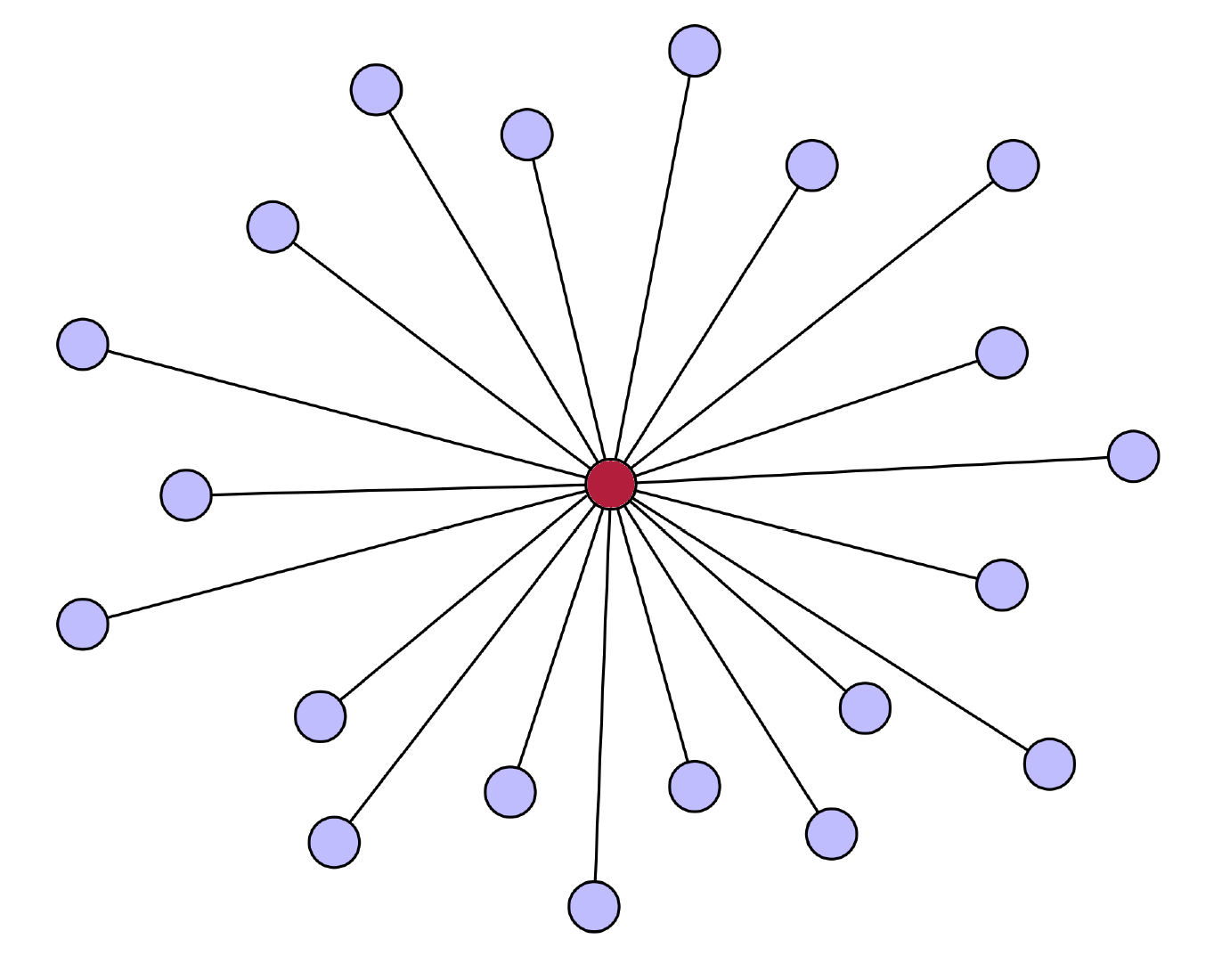}
  \caption{A graph of 3 communities created with the Stochastic Block Model (left) and a star graph (right) colored according to their resulting information access clustering (for $k=3$ and $k=2$, respectively).}
  \label{fig:toy}
\end{figure}  

While these representations are well-defined for any model of information flow, in order to describe examples, run experiments, and discuss computational complexity, we need to fix a mechanism for information spread. As such, for the remainder of the paper, we assume that information is flowing according to the extensively studied independent cascade model~\cite{kempe03maximizing} (also known as the SIR model in epidemiology). In this model, a node exists in one of three states: ready to receive, ready to transmit, and dormant.\footnote{These states are called susceptible, infectious and recovered in the SIR formulation.} Initially, some subset of \emph{seed} nodes possess a bit of information and are ready to transmit, and all other nodes are ready to receive. At each time step, a node that is ready to transmit iterates over all of its neighbors which are ready to receive, and sends the information with probability $\alpha$ (the decision is made independently for each adjacent edge). All such transmissions are imagined to happen simultaneously, after which the transmitting node goes dormant. Since the independent cascade model can be characterized by a single parameter $\alpha$, in what follows we will merely use $s^G_\alpha(v)$ to denote the \infovec. 

Unfortunately, as we show in Section~\ref{sec:hardness}, computing the \infovec under independent cascasde is \#P-hard. In order to get around this intractability for experiments, we estimate the probabilities using Monte Carlo simulations. Specifically, to estimate $p_{ij}$, we run $S$ simulations of independent cascade where $i$ is the only initial seed node and report the fraction of trials in which $v$ successfully received the information; each simulation takes time $O(m)$ since an edge can only transmit information at most once in a given cascade. We observe that in order to compute the full representation, we need to do this with each of the $n$ nodes in $G$ as the initial seed, resulting in a time complexity of $O(Smn)$ and 
requiring space $O(n^2)$.

\subsection{Examples}
To demonstrate the goals and utility of the introduced representation and clustering we discuss both synthetic and real-world motivating examples. 

\paragraph{Synthetic Graphs with Meso-scale Structure} 

Two important identified mesoscale structures are block communities and core-periphery structures \cite{rombach2014core}. Since our proposed clustering is meant to capture structural information access patterns, we expect that in the first case, communities should correspond to clusters (members receive similar information with high access inside their own community and low access from those outside). In the second setting, the information access view should reflect the core-periphery structure, with central nodes having high access to information across the entire network and peripheral nodes all having similarly low access to information from anyone outside the core. 

We now argue that on simple representatives from each of these classes, our approach does recover the desired clusterings 
(each community as a cluster, and a core versus periphery clustering, respectively).

We first consider a synthetic graph consisting of three strong communities of size 100, generated using the stochastic block model~\cite{sbm}. Each edge between members of the same community is present independently with probability 0.3, and 
edges between pairs of nodes in differing communities are present with probability 0.05. The information access representation matrix for this graph has a block structure with access probabilities averaging 0.88 between any two nodes in the same community,  and average access probabilities between 0.62 and 0.67 between communities\footnote{These probabilities were estimated using 10,000 Monte Carlo simulations; the variance across individual nodes for the access between communities $i$ and $j$ was less than .001 in all cases ($1 \leq i \leq j \leq 3$)}. Thus, the signatures for nodes in each community are highly similar (mostly differing at the entries corresponding to the two nodes being compaed), but also easily distinguishable from signatures from any other cluster. We computed an information access clustering (sing the elbow method to determine the appropriate number of clusters $k = 3$), and observe that it does correctly identifies each community as a distinct cluster, as seen in Figure~\ref{fig:toy}.

For core-periphery structure, we use a star graph which consists of a single central node connected to many peripheral nodes, each of which has no other neighbors.  Any distinct pair of nodes in the peripheral set has probability $\alpha^2$ of information transmission between them while any pair involving the central node has probability $\alpha$.  Thus, peripheral node signatures take the form $(1, \alpha^2, \ldots, \alpha^2, \alpha)$ while the central node's signature is $ (\alpha, \alpha, \dots,\alpha, 1)$.
Again, if we select the number of clusters using elbow/silhouette, we find that the central node is in one cluster and all peripheral nodes are together in a second.

These examples demonstrate that the information access clustering we introduce is expressive enough to capture both community-based and core-periphery mesoscale structures.

\paragraph{Real-world example: DBLP co-authorship network}

We begin with an initial motivating experiment on the DBLP co-authorship network of scholars in computing-related disciplines (described in more detail in Section \ref{sec:data}).  What might information access-based social capital or privilege look like in a co-authorship network and how could it be measured?  Given that the co-authorship network is an important form of social network within academia, we posit that one form of privilege in a co-authorship network is access to people in prestigious institutions.  This access could provide field-critical information, i.e. the ability to both receive information \textit{from} and share information \textit{with} people at prestigious institutions, which might then impact faculty hiring (see, e.g., \citep{clauset:etal:2015}). In an undirected network like the DBLP co-authorship network we study, these forms of access can both be modeled as the information access probability between people.

\begin{figure}[t]
    \centering
    \includegraphics[width=2in]{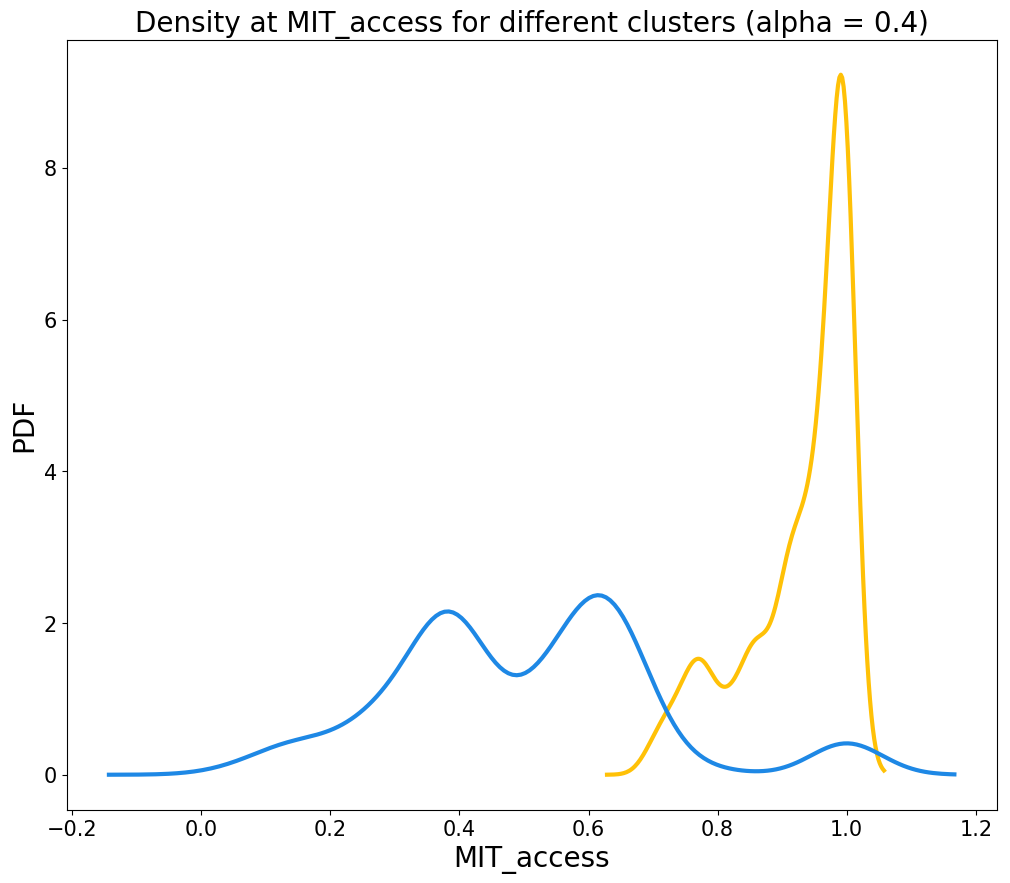}
    \caption{The probability density function showing the probability of access to someone working at MIT through the DBLP co-authorship network for two information access clusters.  The average access probability for the blue cluster is $0.50$ while the average access probability for the yellow cluster is $0.93$.}
    \label{fig:mit}
 \end{figure}

To examine whether the information access clustering groups scholars based on this probability of access to a prestigious institution, we take the case of MIT and examine the probability that an individual in the network is able to reach (or be reached by) someone working there.  Using the same experimental setup described later in this paper and $k=2$ clusters, we created the information access clustering and determined the probability of access to MIT for each individual. We find that one cluster is clearly the privileged cluster; its nodes on average have higher citation count, better job institution rank, and better Ph.D. institution rank. The average access probability across the privileged cluster is about two times that of the non-privileged cluster (see Figure \ref{fig:mit}), and the access to individuals from MIT is also statistically significantly ($p < 10^{-7}$) higher.\footnote{These results hold across $\alpha \in [0.1, 0.7]$.} 
This, coupled with our theoretical considerations, suggests that information access clustering may indeed allow investigation of information privilege or social capital within a network. With this motivation in mind, we next further consider the theoretical foundation of the representation and later consider a more comprehensive set of experiments (see Sections~\ref{sec:data-driv-expl}-\ref{sec:large}).  

\subsection{Spectral analysis}
\label{sec:math-expl}
We now provide some formal insights into structures that information access clustering seeks to find, by
looking more closely at the structure of the representation $R^G_\alpha$ and
how it relates to structures in the underlying graph $G$. For clarity, we drop
the superscript $G$ when the context is clear.

An alternate (and well-known) way to think about the independent cascade
diffusion is as follows. Fix a graph $G$ and define a distribution
$\mathcal{D}_\alpha(G)$ of graphs as follows: delete each edge of $E(G)$
independently with probability $1-\alpha$. Consider any graph $H$ drawn from
$\mathcal{D}_\alpha(G)$. Let $CC(H)$ denote the binary $n \times n$ matrix
indexed by vertices of $G$ such that $CC(H)_{ij}$ is $1$ if $v_i$ and $v_j$ are
in the same connected component of $H$.
It is then straightforward to show that $R_\alpha = \Exp_{H \sim \mathcal{D}_\alpha(G)}[CC(H)]$.
In other words, we can interpret $p_{ij}$ as the probability that $v_i$ and
$v_j$ are in the same connected component of a graph drawn from
$\mathcal{D}_{\alpha}(G)$.

This interpretation also yields a connection to the spectral decomposition of
$G$.
Let $D$ denote the diagonal matrix where $D_{ii} = \text{deg}(v_i)$ and as usual
let $A$ denote the adjacency matrix of $G$. Then the (combinatorial) Laplacian
of $G$ is $L = D - A$.
If $G$ has $\ell$ components, then the nullspace of $L$ is multidimensional, with one
basis being the set of vectors $v_1, \dots v_\ell$ where each $v_i$ is a binary
vector representing the characteristic vector of the $i^{\text{th}}$ connected
component. It then follows that $CC(G) = \sum_{i = 1}^\ell v_i
v_i^\intercal = V_G V_G^\intercal$ where $V_G$ is the basis of the null space in which each of the characteristic
vectors of the connected components is a column.
Summarizing,
$R_\alpha = \Exp_{H \sim \mathcal{D}_\alpha(G)}[V_H V_H^\intercal] $.
This immediately indicates how our approach differs from (say) spectral
clustering, which instead computes the eigendecomposition $L = U \Lambda U^\intercal$, fixes a parameter $d \le n$ and
then forms the representation $S = U_d$ where $U_d$ is an $n\times d$
matrix containing the eigenvectors corresponding to the $d$ smallest
eigenvectors of $L$ (and in particular \emph{ignores} the nullspace).

\subsection{Computation: Hardness \& Simulation}
\label{sec:hardness}
Unfortunately, while the information access representation yields insight into the structure of the
graph, it is intractable to compute exactly under the independent cascade model of information flow. 

\begin{theorem} For an undirected graph $G$ and a fixed probability of transmission $\alpha$, 
  the computation of each entry $p_{ij}$ in the information access representation $R_\alpha^G$ is $\# P$-complete when information is propagated via independent cascade. \label{thm:0}
\end{theorem}

We recall that the entry $p_{ij}$ in the signature is defined to be the probability that node $v_j$ receives information 
seeded at node $v_i$. Shapiro et al proved that computing this quantity is $\# P$-complete~\cite{shapiro2012hardness} under the SIR model with uniform transmission probability -- which, as we have previously observed, is directly equivalent to independent cascade. Their proof is a reduction from the $\# P$-complete \emph{two-terminal network reliability problem}~ \cite{valiant1979complexity}: given a network $G$ of $n$ nodes and $m$ edges where each edge of $G$ is assigned a fixed probability $1-\alpha$ that it disappears from graph (i.e., each edge survives with probability $\alpha$), determine the probability $p'_{ij}$ that in the surviving graph two particular nodes $v_i,v_j$ are connected (have a path between them).

\section{Experiments Part I: Social Capital}
\label{sec:data-driv-expl}

In order to provide validation that information access representations encode information about individuals' social position in a network as it relates to social capital, we examine the information access clustering produced on these representations for a variety of real-world networks. We will consider two main experimental questions:
\begin{enumerate}
\item Does information access clustering allow identification of clusters of individuals in the network that are structurally similar in terms of information access and real-world social capital?
\item How does the information access clustering compare to existing network clustering methods?
\end{enumerate}

We begin with the first question in this section and return to the second in Section \ref{sec:comparisons}.
Code to reproduce all experiments is available at: \url{https://github.com/algofairness/info-access-clusters/releases/tag/paper.1.0}

\subsection{Network datasets}
\label{sec:data}

We choose real-world network datasets that share two fundamental characteristics.  First, we have chosen networks based on domains where information tangibly flows across the network, and where access to that information is clearly a form of privilege within that network.  Second, all networks have an external node attribute that can be used to quantify information access or social capital within the network, which we will refer to as its \emph{external information access measure}.  By external, we mean that this attribute should not be directly encoded in the network structure itself; importance measurements such as node degree do not meet this criterion. Our goal with this criterion is to help answer our first experimental question and determine whether information access clustering, using solely the information access representation, clusters nodes together so that cluster composition is different with respect to the external information access measure. A clustering that separates nodes that are similar in terms of this external characteristic could allow identification of clusters that are advantaged or disadvantaged based on their information access without direct access to a measurement of that advantage, and in general could allow the unsupervised exploration of such external measures given only the information access representation. This would indicate that information about social capital is encoded in the information access representation.
The networks studied range broadly in size -  from $438$ nodes in the Co-sponsorship dataset to $391,642$ nodes in the Google scholar dataset - and experiments are run on the largest (strongly) connected component for each graph. Further considerations are taken into account for dealing with large graphs (see Section \ref{sec:large}). 

\begin{table}[htp]
\begin{center}
\begin{tabular}{c|ccccc} 
Dataset &  
\#Nodes in Largest & \#Edges in Largest & External Information \\
   & Connected Comp.& Connected Comp.& Access Measures  \\
\midrule
DBLP &  
$2,123$ & $7,133$ & citation count, job rank, PhD rank  
 \\

Twitch &  
$7,126$ & $35,324$ & partner, views   \\

Co-sponsorship &
438 & $28,194$ & legislative effectiveness score 
\\

Google Scholar & 
 $391,642$ & $104,647,630$ & citation count, h-index \\

\end{tabular}

\end{center}
\caption{Dataset information.}
\label{tab:datasets}
\end{table}%

\paragraph{DBLP co-authorship}
In the DBLP network, nodes are scholars from \url{https://dblp.uni-trier.de/} and are connected by an edge if the scholars have co-authored a paper. This network contains only scholars who: received their PhD from a university on the Computer Research Association's authoritative Forsythe List of Ph.D.-granting departments in computing-related disciplines in the United States and Canada; had their first assistant professorship at one of these universities; worked at one of these universities in the 2011-2012 academic year; and were hired between 1970 and 2011. This list of scholars, along with metadata about each of them, was compiled by \cite{way:etal:2016}. Co-authors of these scholars were scraped from DBLP in October 2020.  
A small number of scholars were excluded either because their hire date was not known or because their DBLP id was inconsistent with the one recorded by \cite{way:etal:2016}.
The resulting network contains 2,356 nodes and 7,145 edges, with 2,123 nodes and 7,133 edges in the largest connected component.

The external information access measures used for this DBLP network are the number of total citations summed over all papers recorded by Google Scholar for a scholar (node) in the network\footnote{Citation counts were scraped from Google Scholar in Spring 2020.} and the ranks of the Ph.D. and first job institutions of the scholars (as recorded by \cite{way:etal:2016}, ordered so that $1$ is the top rank).  

\paragraph{Twitch}
The Twitch dataset is a social network from Twitch, a video live streaming platform for gamers, where the $7,126$ nodes represent Twitch users who stream in English and $35,324$ undirected edges are mutual friendships between them. The dataset was collected by \cite{rozemberczki2019multiscale} in May of 2018.
Each node in the network is either a ``partner'' or not: a Twitch user becomes qualified to be a ``partner'' by accumulating streaming hours and maintaining more than 75 concurrent viewers on Twitch or having a sizeable audience on other platforms such as YouTube or Twitter. Thus, the external information access measures we use are the number of views for each user and their ``partner'' status.

\paragraph{Congressional co-sponsorship}
The Co-sponsorship data is a directed network indicating bill co-sponsorship for the 114th Congressional sitting in the United States. This dataset was first created by GovTrack and made available by \cite{CosponsorshipNetworkData} in July of 2017 after its removal from the original source.  Based on this data, we created a network in which each node is either a sponsor or original co-sponsor of a bill, and there exists a directed edge from one node to another if the latter originally co-sponsored at least one bill that was sponsored by the former.  The resulting network has $547$ nodes and $33,937$ edges;  the largest strongly connected component of the network consists of $438$ nodes and $28,194$ edges that represent legislators in the House of Representatives.\footnote{There are more than 435 such legislators because of special elections held during the 114th sitting.} 

We use the legislative effectiveness score \cite{LegislativeEffectivenessData} computed for the corresponding sitting  as the network's external information access measure. The score itself quantifies each representative's ability to advance their agenda through the legislative process, from introducing a bill to enacting it as a law, based on 15 weighted indicators \cite{CongressReconsidered}.

\paragraph{Google Scholar}
Finally, we analyze a co-authorship network scraped from Google Scholar.\footnote{\url{https://scholar.google.com}} As in the DBLP co-authorship network, each node represents a scholar and scholars are connected by an edge if they have co-authored a paper. The data were originally collected in May of 2015 by \cite{chen2017google}.\footnote{For the original paper, Chen et al. eliminated 409,961 nodes for having unreliable data. We use the full dataset, which includes any inaccuracies present in Google Scholar.} 
They used
a web crawler to search each letter in the English alphabet using the Google Scholar ``search authors'' feature
and parsed those results to gather the 
scholars' metadata and publication records. 
We added edges between these author nodes if 
two authors had the same paper listed in their set of published papers. This resulted in a total of $812,351$ nodes and $262,933,633$ edges, with $391,642$ nodes and $104,647,630$ edges in the largest connected component.   
We use total citation count as an external measure for influence (these citation counts are available for all nodes)
as well as $h$-index (as recorded by Google Scholar). 

We include the Google Scholar network in our analysis mostly due to its large size, the implications of which we will discuss further in Section \ref{sec:large}. For now, we focus on the setup and analysis of the other networks.

\subsection{Experimental setup}
\label{sec:setup}

\paragraph{Calculating information access signatures}
We calculate the information access vector of node $j$, ~ $s^{G}_{\alpha}(v_j) = (p_{1j},\dots,p_{ij},\dots,p_{nj})$, and estimate each $p_{ij}$ by setting node $i$ as the seed and running $S = 10,000$ independent cascade simulations. We then use the fraction of simulations in which node $j$ received the information to estimate its information access probability.

\begin{figure}[htbp]
\begin{center}
\begin{tabular}{ccc}
~~~$\alpha = 0.1$ & ~~~$\alpha = 0.3$ & ~~~$\alpha = 0.5$  \\

 \includegraphics[align=c, width=0.3\linewidth]{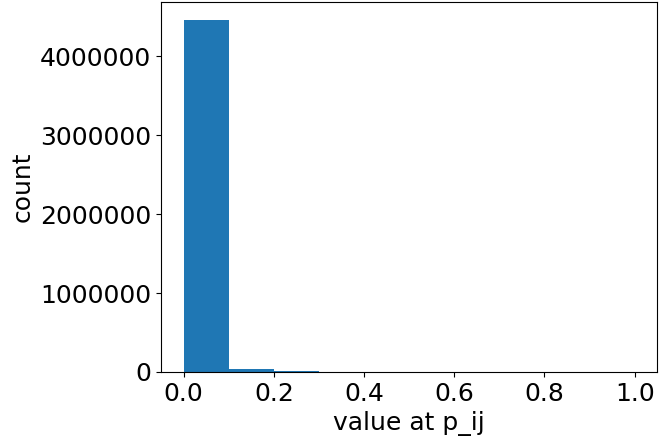} &
 \includegraphics[align=c, width=0.3\linewidth]{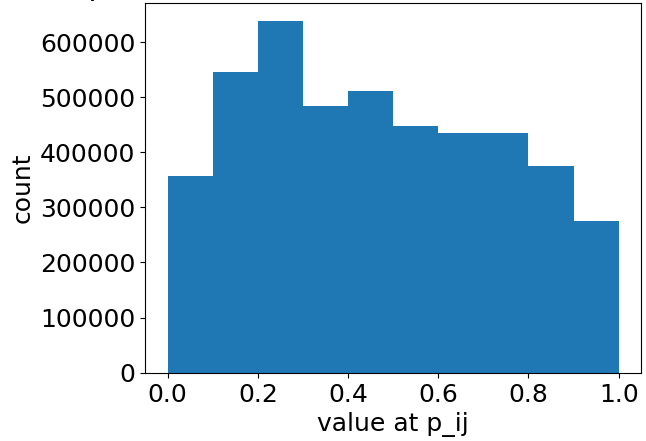} &
 \includegraphics[align=c, width=0.3\linewidth]{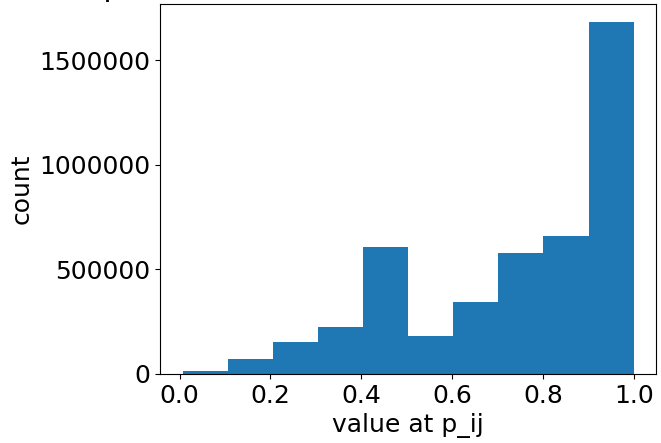} 
\end{tabular}
\end{center}
\caption{Histograms of probability values for the DBLP network and $\alpha = \{0.1, 0.3, 0.5\}$.}
\label{fig:dblp_hist}
\end{figure}

\paragraph{Choosing $\alpha$}  The information access representation, and thus the clustering, are dependent on the choice of $\alpha$.  Since $\alpha$ represents the probability that information is transmitted along an edge, the choice of $\alpha$ is domain-dependent.  We will thus present our results across a range of $\alpha$ values.  However, choosing an appropriate range for $\alpha$ (which is also domain-dependent) is important in order to find $p_{ij}$ values that are not zero or one: at the extremes based on a specific domain, most $p_{ij}$ values in the information access signatures will be zero (indicating that the $\alpha$ value was too small to allow for information transmission across the network) or most will be one (indicating that the $\alpha$ value was large enough that essentially all nodes received information from essentially all other nodes).  \emph{Thus, information access signatures are most interesting when a range of probability values $p_{ij}$ are included in the resulting information access signatures.} To determine those ranges, we considered the histograms for each dataset (a selection of these are shown in Figure \ref{fig:dblp_hist} with the full set of histograms in Appendix \ref{apx:alpha}.
We choose $\alpha \in [0.1, 0.7]$ at increments of $0.1$ for the DBLP and Twitch datasets and $\alpha \in [0.01, 0.07]$ at increments of $0.01$ for the Co-sponsorship dataset since these ranges included a wide spread of histograms, including those with essentially all probabilities close to $0$ and also where essentially all probabilities are close to $1$.  Larger values ranging up to the maximum of $1$ were considered, but above the chosen range essentially all $p_{ij}$ values were $1$. In order to further assess whether the examined $\alpha$ values would provide enough range and granularity, the resulting clusterings were also examined for consistency across $\alpha$ values (see Appendix \ref{apx:clustering_consistency}).

\begin{figure*}[h]

\begin{center}
\textbf{Elbow Method}\\
\begin{tabular}{ccc}
~~~$\alpha = 0.1$ & ~~~$\alpha = 0.3$ & ~~~$\alpha = 0.5$  \\
 \includegraphics[align=c,height=1.1in]{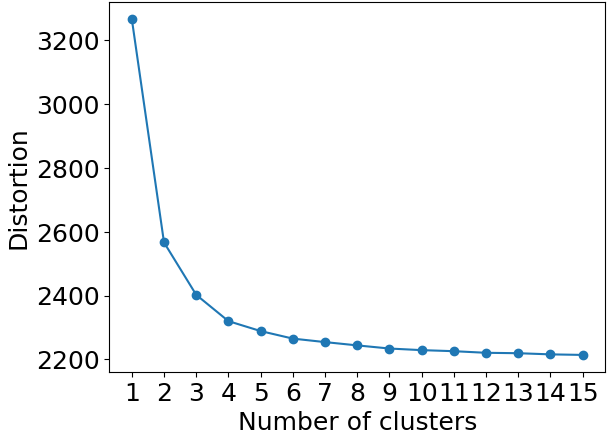} &
 \includegraphics[align=c,height=1.1in]{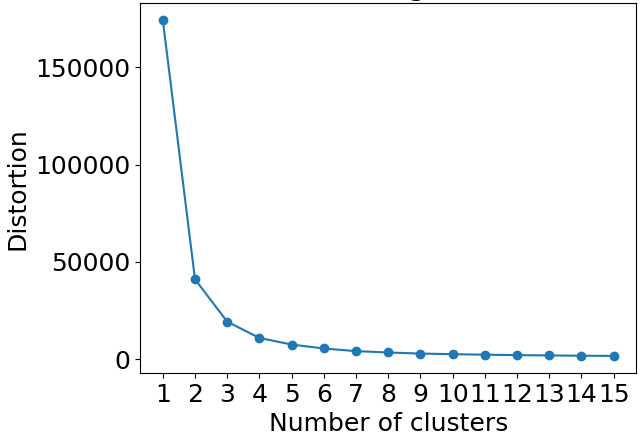} &
 \includegraphics[align=c,height=1.1in]{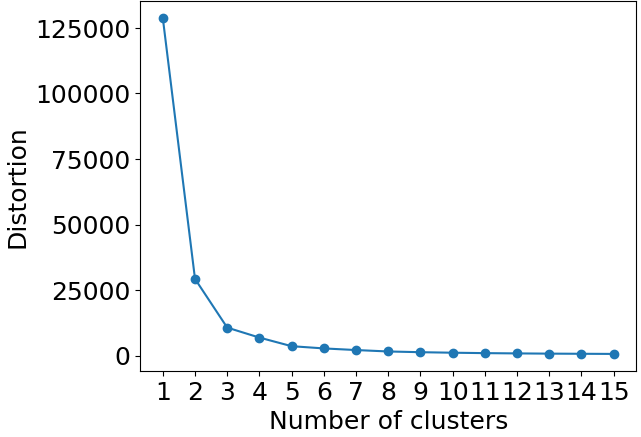}
\end{tabular}

~\\
\textbf{Silhouette Values}\\
\begin{tabular}{c|ccccccccc}
& \multicolumn{8}{c}{$k$} \\
$\alpha$ & 2 & 3 & 4 & 5 & 6 & 7 & 8 & 9 & 10\\
\hline
0.1 & \textbf{0.279} & 0.140 & 0.072 & 0.053 & 0.035 & 0.033 & 0.023 & 0.024 & 0.016 \\
0.3 & \textbf{0.658} & 0.605 & 0.578 & 0.558 & 0.551 & 0.532 & 0.517 & 0.501 & 0.485\\
0.5 & \textbf{0.753} & 0.740 & 0.717 & 0.723 & 0.676 & 0.657 & 0.644 & 0.640 & 0.631
\end{tabular}
\end{center}
\caption{Elbow method charts and silhouette values for the DBLP dataset demonstrating that $k=2$ for information access clustering on this dataset.}
\label{fig:choose-k_dblp}
\end{figure*}

\paragraph{Choosing $k$} Each clustering is dependent on the value of the number $k$ of clusters  chosen.  This also depends on the domain / network.  We use the silhouette value \cite{rousseeuw1987silhouettes} and elbow method to determine the appropriate values of $k$ for each dataset, looking for consistency across those two methods (see Figure \ref{fig:choose-k_dblp} for selected $\alpha$ value results on the DBLP dataset and Appendix \ref{apx:choose-k} for full results).  The DBLP and Co-sponsorship datasets have matching silhouette values and elbow method plots across all considered $\alpha$ values indicating that $k=2$.  The Twitch dataset elbow method plots indicate that $k=2$ for all $\alpha$ values, while the silhouette values indicate that $k=2$ for low values of $\alpha$ and has some slightly larger silhouette values for other $k$ values at higher $\alpha$ values.  Since the elbow method plots consistently indicate that $k=2$, the silhouette value differences are small, and the Twitch dataset has a known domain-specific $k=2$ value from the ``partner" flag described above, we also use $k=2$ for the Twitch dataset. We note here the coincidence that $k$ was found to be 2 for all of our networks; though we do not explore in this paper potential reasons behind and implications of this.

\subsection{Clusters and external information access measures}

\begin{table}[htbp]
\begin{center}
\begin{tabular}{c|ccc|cc|c } 
 & \multicolumn{3}{c}{DBLP} &\multicolumn{2}{c}{Twitch} & \multicolumn{1}{c}{Co-sponsorship}   \\
 & Citation Count & PhD Rank & Job Rank &   
 Partner & $\log($ Views $)$ & Legislative \\ 
Clustering &  & && &  & Effectiveness\\ %
\midrule
\textbf{Info Access} & $\bm{< 10^{-7}}$ & $\bm{< 10^{-7}}$ & $\bm{< 10^{-7}}$ 
&  $\bm{< 10^{-7}}$ & $\bm{< 10^{-7}}$ & $\bm{< 10^{-7}}$ \\
\midrule
Spectral & 0.762 & \textbf{2.31e-3} & \textbf{1.30e-3} & 
1 & 0.857 & \textbf{0.050} \\ 
Fluid Comm. & \textbf{0.0097} & $\bm{< 10^{-7}}$ & $\bm{< 10^{-7}}$  
&    $\bm{< 10^{-7}}$ & $\bm{< 10^{-7}}$ &     $\bm{< 10^{-7}}$ \\ 
Louvain & 0.164 & 0.226 & \textbf{0.013}  
&    1 & 0.224 &   $\bm{< 10^{-7}}$ \\ 
Role2Vec & $\bm{< 10^{-7}}$ & $\bm{< 10^{-7}}$ & $\bm{< 10^{-7}}$  
&   \textbf{0.007} & $\bm{< 10^{-7}}$ &    $\bm{< 10^{-7}}$\\
Core-Periphery & $\bm{< 10^{-7}}$ & $\bm{< 10^{-7}}$ & $\bm{< 10^{-7}}$ 
&    $\bm{< 10^{-7}}$ & $\bm{< 10^{-7}}$ &   0.472 \\  
\end{tabular}

\end{center}
\caption{Kruskal-Wallis and Fisher Exact test (for the partner  
categorical data) $p$-values testing to see if the distributions of the external information access measures are different across clusters.  Minimum $p$-values across all tested $\alpha$ values are given for the information access clusters, shown with an applied Bonferroni correction of $10$. 
The fluid communities algorithm includes some randomness, so values shown are the minimum $p$-value for that attribute over $10$ runs with a Bonferroni correction of $10$ applied.  Resulting $p$-values less than or equal to $0.05$ are shown in bold.  Very small $p$-values are shown as $< 10^{-7}$.}
\label{tab:results_summ}
\end{table}

We hypothesize that since information access also controls and/or is correlated with people's access to other resources, this implies that certain external measures such as academic productivity and privileged status in an affiliate program should correlate with access to information.
We thus further hypothesize that \emph{external information access measures will be different between different clusters}, indicating a difference in the social capital of each cluster and the structural similarity within clusters in terms of network position.
In this section, we evaluate this hypothesis for the DBLP, Twitch, and congressional co-sponsorship networks.
Specifically, each cluster induces a distribution of external information access measures.
If these distributions significantly differ from one another, this provides evidence consistent with our hypothesis.
We use the Kruskal-Wallis and the Fisher exact tests to check the similarity of distributions.
There are 10 total experiments for each dataset (by varying $\alpha$ across a range of plausible values), and so we apply a Bonferroni correction factor of 10 to all reported results\footnote{This correction is 10 and not 7 since tests not included in the paper were originally run on a larger range of $\alpha$ values; we exclude three of these runs from the paper because of analysis described in Section~\ref{sec:setup}.}
See Table~\ref{tab:results_summ} for all results.

\begin{figure}
\begin{center}
\begin{tabular}{c|ccc}
$\alpha$ & ~~~~~\textbf{Citation Count} & ~~~~~~~~~~\textbf{Ph.D. Rank} & ~~~~~~~~\textbf{Job Rank}\\
0.1 & \multicolumn{3}{c}{\includegraphics[align=c,height=1.2in]{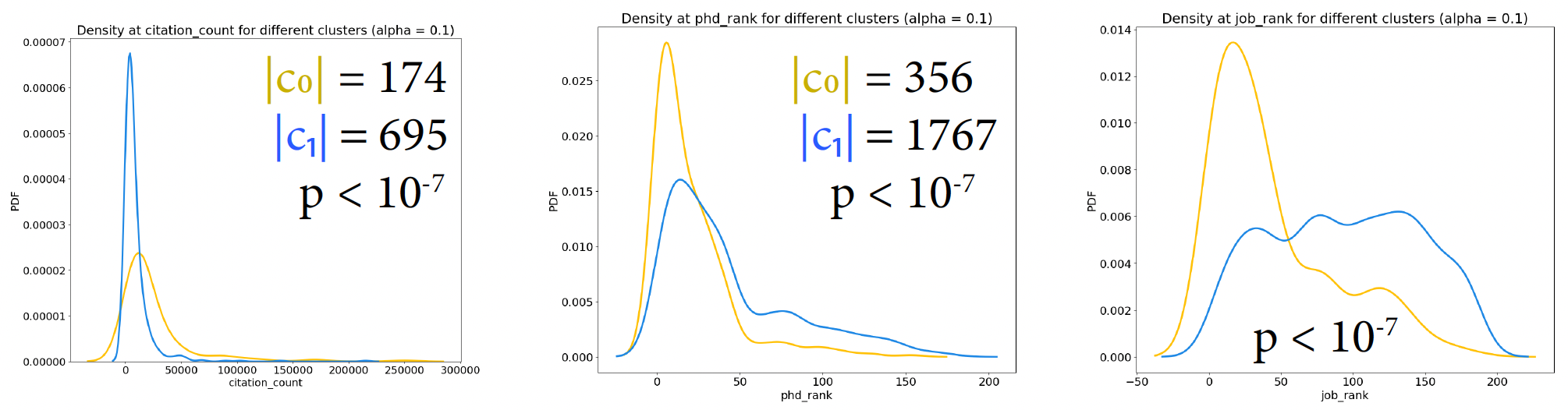}}\\
0.3 & \multicolumn{3}{c}{\includegraphics[align=c,height=1.2in]{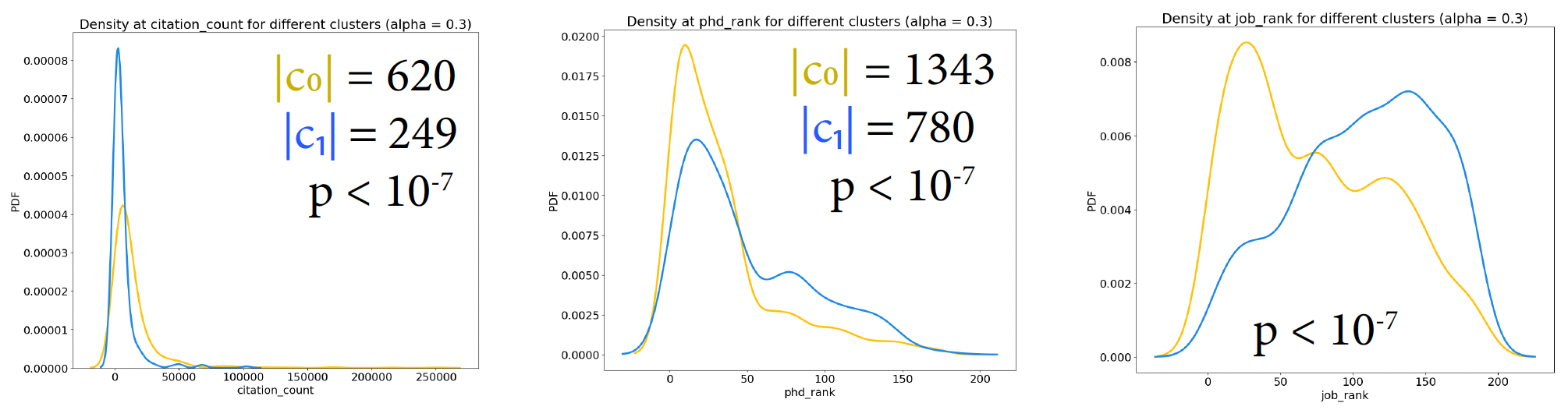}}\\
0.5 & \multicolumn{3}{c}{\includegraphics[align=c,height=1.2in]{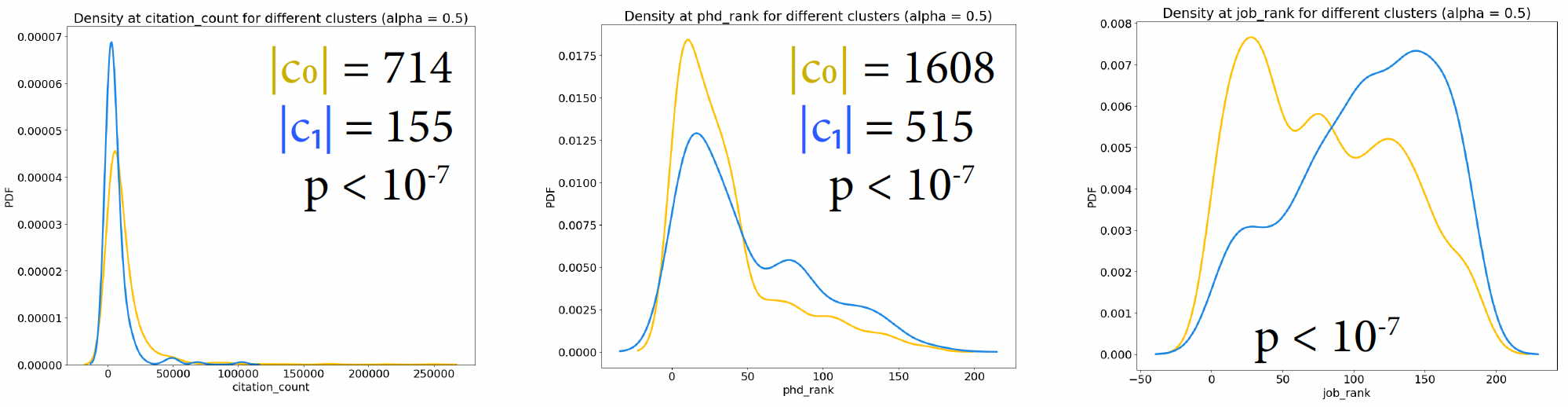}}\\
\end{tabular}
\end{center}
\caption{Information access clusterings on the DBLP dataset for $\alpha = \{0.1, 0.3, 0.5\}$. The probability density function (PDF) per cluster is shown with respect to the indicated external information access measure.}
\label{fig:dblp_prob_density}
\end{figure}

Looking at the probability density functions per cluster with respect to the external information measure allows us to better understand cluster composition. These results are shown for select $\alpha$ values on the DBLP dataset in Figure \ref{fig:dblp_prob_density} with full results in Appendix \ref{apx:prob_density}. We find that one of the DBLP clusters has more researchers with higher citation counts and better job and PhD institution ranks than the other, one of the Twitch clusters has more individuals with higher views and partner status, and one of the Co-sponsorship clusters has more individuals with higher legislative effectiveness scores.
These clusters contain the more influential individuals in the network, with higher access to information and better ability to spread that information or reach important individuals in the network.
Overall, these results thus show that information access clustering is able to create clusters that separate nodes by external information access measure across all three datasets.
This evidence is consistent with our hypothesis and indicates that \emph{information access representations encode information relevant to real-world social capital}.

\section{Experiments Part II: Comparisons} 
\label{sec:comparisons}

We now consider our second question: whether other clustering methods create clusterings that separate nodes based on external information access measures, and whether information access clustering is consistently similar to one or more of these other methods.  I.e. do existing methods already identify emergent network privilege by creating essentially the same clusters as those chosen by information access clustering?  We will show that the answer is no.

\subsection{Comparison methods}
The clustering method we introduce does not cleanly belong to the major graph clustering method groups (see Section \ref{sec:related-work}), so we compare against one method from each of the following groups:

\emph{Spectral methods.} The well-known spectral clustering technique uses the eigenvectors of the graph Laplacian as node representation, which it then uses to cluster the vertices.
We compare to \texttt{sklearn}'s \texttt{SpectralClustering} \cite{scikit-learn} with default parameters: as many eigenvectors for the representation as the number of clusters, and a normalized Laplacian.

\emph{Community detection.} Community detection methods are concerned with finding dense subgraphs based on direct connections between individuals.
We compare to two such methods.
The Louvain method attempts to optimize the modularity of a cluster (the relative density of links internal to a cluster versus those connecting outside the community~\cite{blondel2008fast}\footnote{\url{https://python-louvain.readthedocs.io/en/latest/}}). The fluid communities method is based on the idea of propagation of a fluid within a community, reaching an equilibrium with neighboring communities in the network from randomly chosen seed nodes \cite{pares2017fluid}.\footnote{\url{https://networkx.org/documentation/stable/reference/algorithms/generated/networkx.algorithms.community.asyn_fluid.asyn_fluidc.html}}

\emph{Role detection.}
Role detection algorithms cluster nodes together if they serve similar roles in the network.
Nodes are often based on locally defined motifs, e.g., nodes that bridge two otherwise disconnected components.
We compare to \texttt{role2vec}~\cite{ahmed2018learning}\footnote{\url{https://github.com/benedekrozemberczki/role2vec}}.
This method creates a vector representation of the network by considering random walks that balance an outward exploratory focus with returning to the start of the walk.
Role2vec additionally takes an input motif to help guide these walks; we use an identity matrix motif to focus on individual nodes, since our other comparisons also assume no additional information.

\emph{Core-periphery clustering.}
Finally, core-periphery methods assume a network has a \emph{core} group of nodes that are densely connected and a \emph{periphery} that is well-connected to the core but not well connected within the periphery; a classic example is a star graph.
We compare to Rombach et al.'s method \cite{rombach2014core}.\footnote{\url{https://github.com/skojaku/core-periphery-detection/blob/master/cpnet/Rombach.py}}
This approach learns a mapping from nodes to the $[0,1]$ interval with larger value signifying greater indication of being a core vertex.
The mapping is constrained to agree with the core-periphery characterization: two periphery vertices should not be connected to each other, and two core vertices should. 

\subsection{Assessing clustering similarity}
Using the same experimental setup described in the previous section, we find (see Table~\ref{tab:results_summ}) that many of these comparison methods create clusters that are separated in terms of their external information access measure.\footnote{In some cases where they don't, the clusterings produced with $k=2$ on some datasets have a very small number of nodes in one of the clusters.
In these cases, sample sizes are so small that they provide no statistical evidence of the difference between cluster distributions.
We observe this for spectral clustering and the Louvain method, on both the DBLP and Twitch datasets.}
Overall, both community detection methods succeed at separating clusters according to these access measures for the co-sponsorship data, but Louvain fails on one more external access measures for the DBLP and Twitch data.
Fluid community results differ drastically depending on the random initialization of the clusters; some resulting clusterings successfully distinguish information access measures while others do not.
The core-periphery method, on the other hand, fails to separate based on the legislative effectiveness measure for the co-sponsorship data while it succeeds across all measures for the DBLP and Twitch data.
The role detection method \emph{does} successfully create clusters with different information access distributions across all datasets and access measures, as does information access clustering.

\newcommand{\smallari}{$\sim 0$}
\begin{table}[htb]
\begin{center}

\begin{tabular}{c|ccccc}
$\alpha$ & Spectral Clustering & Fluid Communities & Louvain & Role2Vec & Core-Periphery \\
\hline
0.1 & \smallari & \smallari & \smallari & \smallari & 0.73 \\
0.3 & \smallari & \smallari & \smallari & 0.49 & \smallari \\
0.5 & \smallari & \smallari & 0.01 & 0.38 & \smallari
\end{tabular}
\end{center}

\caption{For the DBLP dataset and each $\alpha$-parameterized clustering, the above table gives the adjusted rand index indicating the difference between the resulting information access clustering and the indicated clustering method.  The adjusted rand index is $0$ when two clusterings do not agree on any pair of points. For legibility, adjusted rand index values less than $0.01$ are shown as $\sim 0$. Fluid communities method given values are the mean across $10$ random runs.}

\label{tab:dblp_ari_comparisons}
\end{table}%

To further assess the similarity of these clustering methods, we directly compared the resulting clusterings to determine whether nodes were assigned to the same groupings across clusterings (see Table~\ref{tab:dblp_ari_comparisons} for selected results on the DBLP dataset and Appendix \ref{sec:apx-comparisons} for full results). No method created clusterings that were similar to information access clusterings across all datasets.

\begin{table}[htb]
\begin{center}

  \begin{tabular}{%
    >{\raggedright}p{0.18\columnwidth}
    |>{\centering}p{0.1\columnwidth}>{\centering}p{0.1\columnwidth}
     |>{\centering}p{0.1\columnwidth}>{\centering}p{0.1\columnwidth}
     |>{\centering}p{0.02\columnwidth}>{\centering\arraybackslash}p{0.02\columnwidth}} \toprule
 Inf. Acc. $\alpha$ & \multicolumn{2}{c}{DBLP} & \multicolumn{2}{c}{Twitch} & \multicolumn{2}{c}{Co-sp.} \\ \midrule
$0.1$ or $0.01$ & \textbf{1} & \textbf{72} & \textbf{1} & \textbf{2461}&  \textbf{1} &  \textbf{1}\\
$0.2$ or $0.02$ & \textbf{1} & \textbf{374} & \textbf{1} & \textbf{2674} &  \textbf{1} &  \textbf{1} \\
$0.3$  or $0.03$& \textbf{1} & \textbf{455} & \textbf{1} & \textbf{2169} &  \textbf{1} &  \textbf{1} \\
$0.4$ or $0.04$ & \textbf{1} & \textbf{422} & \textbf{1} & \textbf{2012}&  \textbf{1} &  \textbf{1} \\
$0.5$ or $0.05$ & \textbf{1} & \textbf{414} & \textbf{1} & \textbf{2015} & 1 & 1 \\
$0.6$  or $0.06$& \textbf{1} & \textbf{233} & \textbf{1} & \textbf{1212}& 1 & 11 \\
$0.7$  or $0.07$& \textbf{1} & \textbf{224} & \textbf{1} & \textbf{1173}& 1 & 6 \\
\midrule
Spectral & \textbf{1}& \textbf{1} &1 & 5 & \textbf{1} & \textbf{1} \\
Fluid C$^*$. & $\bm{3}$ & $\bm{4}$ & $\bm{45}$ & $\bm{48}$ & \textbf{1} & \textbf{1} \\    
Louvain & \textbf{1} & \textbf{1}&1 & 1 & \textbf{1} & \textbf{1} \\
Role2Vec & \textbf{4} & \textbf{299} & \textbf{253} & \textbf{674} & \textbf{1} & \textbf{1}\\
 Core-Peri. & \textbf{1} & \textbf{102} & \textbf{1} &  \textbf{1756} & 1 & 1\\ \bottomrule
\end{tabular}

\end{center}

\caption{Number of connected components per cluster by clustering type and
  dataset.  For clustering methods that distinguish between more or less
  influential clusters along at least one external information access measure
  (see Table~\ref{tab:results_summ}), the more influential cluster is listed
  first and the entries are given in bold.  Fluid communities numbers are the
  mean values over different seeds.}

\label{tab:connected_components}
\end{table}%

To further investigate and understand the similarities and differences between these methods, we determined the number of connected components per cluster for each of these methods (see Table~\ref{tab:connected_components}).  This allows us to see that information access clustering (for $k=2$) creates one cluster with a single connected component and another cluster that is a collection of disconnected components on the DBLP and Twitch datasets.  In all cases where these clusters distinguish the external access measure, the cluster with the single connected component also contains more of the higher information access (privileged) nodes.  In this way, information access clustering is similar to the core-periphery methods that create a single core or influential cluster that's connected and another cluster to represent the periphery.  Again, on the co-sponsorship dataset we see that information access clustering is more similar to community detection methods, with both clusterings creating two single component clusters (for all significant $\alpha$ values).  Finally, we see that role2vec does not follow the same core-periphery pattern as information access clustering for the DBLP and Twitch data, creating two clusters with multiple components.
Thus, information access clustering appears to act differently from 
all examined methods and is able to identify privileged clusters across both core-periphery and community structured networks.

\section{Experiments Part III: Large networks}
\label{sec:large}
\label{sec:seed-selection-large}

The previous experiments considered networks that have at most $7,126$ nodes.  The Google Scholar network, however, has $391,642$ nodes and $104,647,630$ edges in its largest strongly connected component which makes calculating the full information access signature for each node prohibitively slow.  In this section, we assess selection of a smaller signature based on a number of possible seed selection methods. We first use the DBLP co-authorship network to assess the extent to which the resulting information access clustering matches that performed on a full representation, and then consider the Google Scholar co-authorship network.

\subsection{Computing information access representations for large graphs} 
\label{sec:theory-large}
Computing the information access signature of a node $v_j \in V$ demands determining its probability of receiving information from every other node in $V$. This becomes computationally intensive for large networks. We therefore consider a smaller signature (abusing notation slightly), $s_\alpha^I(v_j) = (p_{1j},\dots  p_{ij}, \dots p_{bj})$ for all $v_i \in I$, where $I \subset V$ and $|I| = b$, i.e. instead of considering the signature representing each other node as a possible source of information, we choose a subset of \emph{seed} nodes and create each information access signature only in terms of those nodes.  

Based on these smaller signatures, we determine the representation $R^I_\alpha = \{ s_\alpha^I(v_i) | v_i \in V \}$ and treat this as the network representation. This creates a matrix representation of size $n$ by $b$, i.e., this smaller information access signature is created for each node. In cases where the graph is too large for local memory, it's not possible to update all signatures simultaneously. However, the probabilities $\{p_{i1},\dots p_{ij}, \dots p_{in}\}$ for each $v_i \in I$ can be computed independently based on a single information propagation experiment and then merged to create the representation. Since we use the independent cascade method for our experiments, we do this by simulating an independent cascade starting from the single node $v_i$ and determining reachability for all nodes in the graph. As described in Section \ref{sec:setup}, these are run 10,000 times and averaged to determine the probability estimates. Usefully for large networks, computing the probabilities one seed at a time in this way will likely mean that a large portion of the graph remains unreached by the independent cascade, and thus does not need to be stored.

Ideally, the smaller access signatures resulting from the seed set $I$ would be similar to those resulting from the full network, and would also result in similar clusterings. We leave a full theoretical analysis of how to select such an $I$ as an open problem; and limit ourselves to four heuristic approaches to picking $I$. These seed selection heuristics are designed to take advantage of intuition that choosing highly connected or otherwise central nodes as seeds will allow most nodes in the graph to be reached and thus allow those, potentially peripheral, nodes to have non-zero terms in the resulting signature. Thus we will evaluate three deterministic strategies that rank nodes according to a centrality criterion and pick the top $b$ of these. The three criteria we use are (i) (global) PageRank, (ii) betweenness centrality, and (iii) node degree (also known as degree centrality). Additionally, we examine the efficacy of choosing $b$ seeds uniformly at random.

With these modifications, the time to compute the full representation can be reduced from $O(Smn)$ to $O(Smb)$. For each seed we compute the signature independently (which takes $O(Sm)$ by doing $S$ repetitions of breadth first search from that seed). Therefore, computing this smaller representation takes $O(C) + O(S \cdot m \cdot b)$ time and $O(bn)$ space where $S$ is the number of simulations we run to estimate the probabilities, $C$ is the time it takes the chosen selection strategy to select all seeds, and $n$ and $m$ are, respectively, the number of nodes and edges in the graph. In the case of the random seed selection strategy, this is simply $O(Smb)$.

\subsection{Experimental results}

Given the goal of using the seed set $I$ to compute a smaller information access representation that still generates clusterings that are similar to those from the full representation, we next consider the DBLP network. The DBLP network is small enough to compute the full representation, so we compare its information access clustering on the smaller signatures and on the full representation, using the adjusted rand index (adjusted rand index values are $1$ when two clusterings agree on all pairs of points). We consider the four selection strategies for choosing the set of seed nodes (see Section \ref{sec:theory-large}) and run the information access clustering on the resulting smaller representation.

We find that clusterings that are very similar to those developed via the full representation can be created even with a small number of selected seeds.  We tested seed sets of size 5 to 70 at increments of 5 and found that for $\alpha \in [0.2, 0.7]$, the adjusted rand index was at least $0.99$ for the clusterings resulting from all centrality-based seed selection strategies; randomly selected seeds generated clusterings with adjusted rand indexes at least $0.97$ for the same parameters, so the random strategy was also highly successful. While the DBLP dataset has $2,123$ nodes, smaller signatures with only 10 or more seeds create clusterings with an adjusted rand index of at least $0.97$ when compared with information access clustering for values of $\alpha > 0.1$ over all selection strategies.  We find that all three centrality-based selection strategies perform better than random selection, but random selection also performs well (see results for $\alpha = 0.1$ and $\alpha = 0.2$ in Figure \ref{fig:large-ari} and full results in Appendix Table \ref{fig:sampling_aris}), creating clusterings that are essentially the same as those from the full representation. Since random sampling is faster than the other strategies, we adopt that strategy for experiments on the larger Google Scholar network.

\begin{figure}[htbp]
    \centering
    \includegraphics[width=0.45\linewidth]{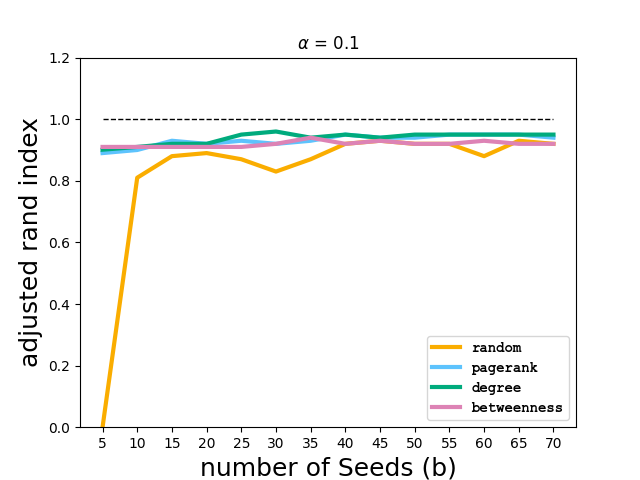} 
     \includegraphics[width=0.45\linewidth]{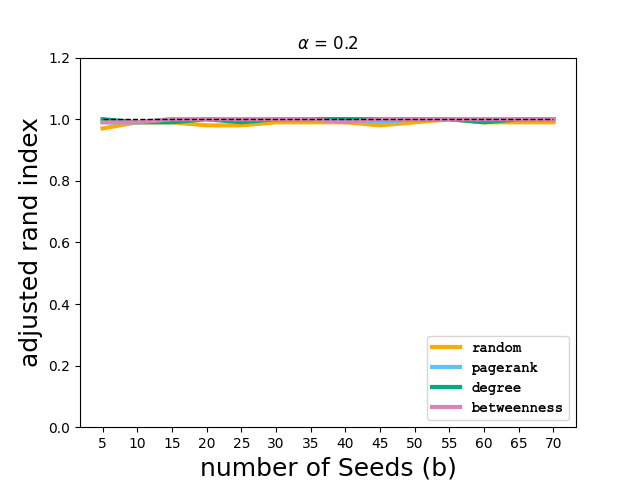} 
    \caption{Results for four seed selection strategies to create smaller information access representations are shown for the DBLP dataset. Adjusted rand indices of the resulting clusterings when compared to the full representation clustering are shown for $\alpha = 0.1$ (left) and $\alpha = 0.2$ (right) with seed set sizes at increments of 5 from 5 to 70. The results for $\alpha > 0.2$ are essentially the same as those for $\alpha = 0.2$ and are given in Appendix Table \ref{fig:sampling_aris}. Adjusted rand indices are $1$ when two clusterings are the same.}
    \label{fig:large-ari}
\end{figure}

Recall from Section \ref{sec:data} that the Google Scholar network has $391,642$ nodes. We choose the size of $I$ to be $b = 632$ since that is the square root of the total number of nodes in the network. This simple heuristic for sample size selection was found to be an effective size choice in the previous section on the DBLP dataset which had $2,123$ nodes (where a sample of $46$ nodes produced adjusted rand indices of at least $0.99$ on most $\alpha$ values indicating essentially the same clustering as would have been calculated using the full information access representation). Using the same experimental setup as described in Section \ref{sec:setup}, we considered the histograms showing the prevalence of varying $p_{ij}$ values within the information access signatures varying over $\alpha \in \{ 0.01, 0.03, 0.05, 0.4 \}$. The resulting range of probabilities includes $\alpha$ values that generate low, medium, and high spreading scenarios, as desired. The silhouette values and elbow method plots were generated for the chosen $\alpha$ values, and all methods determined that $k=2$. Full results of these experimental setup experiments can be found in Appendix \ref{apx:gscholar}.

Information access signatures were calculated on a shared computing cluster, with run time depending on the availability of node resources as well as $\alpha$. On average the run time for a single seed node was roughly $5$ minutes, for a total of approximately $2$ days to generate the information access representation for a specific $\alpha$ given the seed size of $632$.
As before, we investigate the correlation between clusterings and the Google Scholar network's external information access measure. 
We use the Kruskal-Wallis test   
and find evidence ($p < 10^{-7}$) across all $\alpha$ values that the distribution of citation counts and $h$-index values is different between clusters in the information access clusterings.

\section{Discussion and Conclusion}

In this paper, we introduced a representation of individuals in a network based on their information access, creating a mathematical representation of social capital in a network.  Given this social capital representation, we introduced a technique to cluster individuals based on network privilege, as indicated by information access. 
We provided both social theory and mathematical formal insights into what the
underlying representation captures, showed that calculating this representation is \#P-hard, and provided practical heuristics for its calculation on large graphs.
Using real-world data, we validated the encoding of social capital information in the introduced information access representations, demonstrating that clustering on this representation effectively separates individuals based on external measures of information access.
We showed experimentally that these clusterings are
different than existing community-based, role detection, and core-periphery methods.

While we chose these information access measures to be purposefully \emph{external} to the network, given that the information access clustering was able to create clusters that distinguish these measures using only information encoded in the network, these measures are not truly external.  This brings up an interesting question of causality -- did the network structure impact these ``external" measures or did the measures of social capital impact the structure of the network?  
We leave this interesting, potentially domain-specific, question for future work.

\newpage

\bibliographystyle{ACM-Reference-Format}
\balance
\bibliography{access,refs}

\newpage


\appendix

\section{Experimental setup details}

\subsection{Choosing $\alpha$}
\label{apx:alpha}

\begin{figure*}[h!]
\begin{center}
\begin{tabular}{cccc}
\begin{tabular}{c|c}
$\alpha$ & DBLP \\
$0.1$ & \includegraphics[align=c, height=1in]{figs/dblp/probability_histograms/dblp_1_probability_composition.png} \\
$0.2$ & \includegraphics[align=c, height=1in]{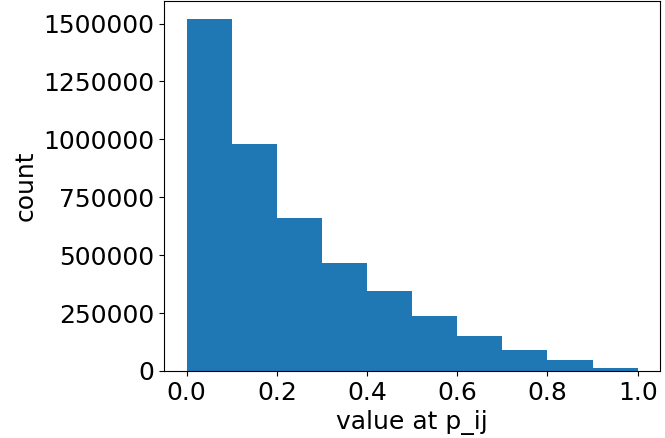} \\
$0.3$ & \includegraphics[align=c, height=1in]{figs/dblp/probability_histograms/dblp_3_probability_composition.png}\\
$0.4$ & \includegraphics[align=c, height=1in]{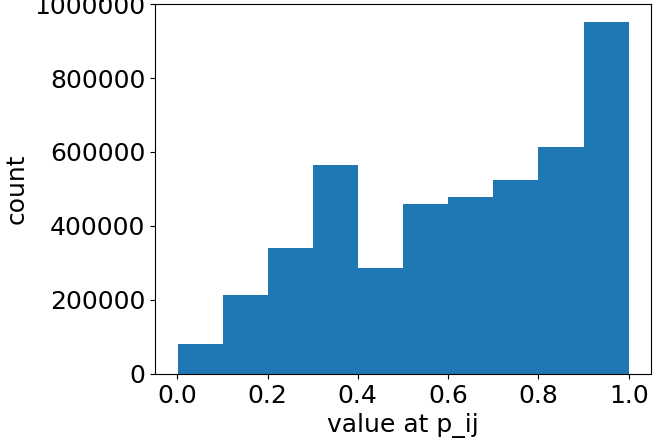} \\
$0.5$ & \includegraphics[align=c, height=1in]{figs/dblp/probability_histograms/dblp_5_probability_composition.png} \\
$0.6$ & \includegraphics[align=c, height=1in]{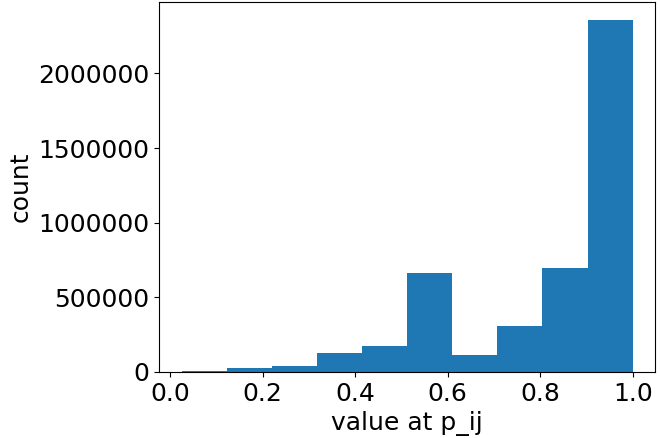}  \\
$0.7$ & \includegraphics[align=c, height=1in]{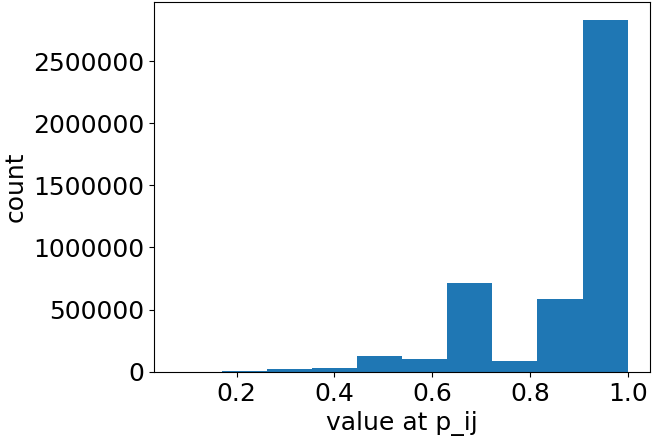}  
\end{tabular}
&
\begin{tabular}{c|c}
$\alpha$ & Twitch\\
$0.1$ & \includegraphics[align=c, height=1in]{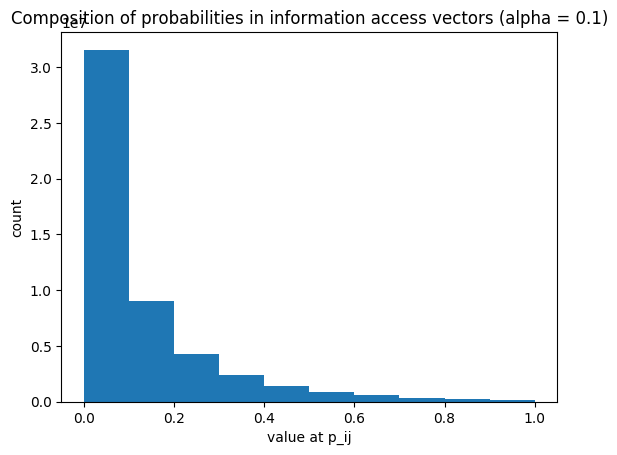} \\
$0.2$ & \includegraphics[align=c, height=1in]{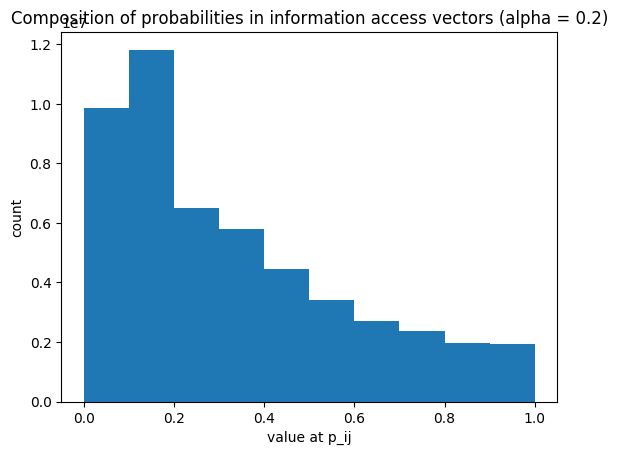} \\
$0.3$ & \includegraphics[align=c, height=1in]{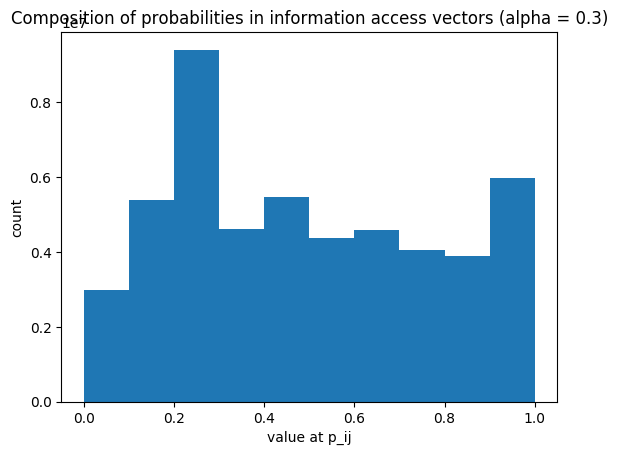} \\
$0.4$ & \includegraphics[align=c, height=1in]{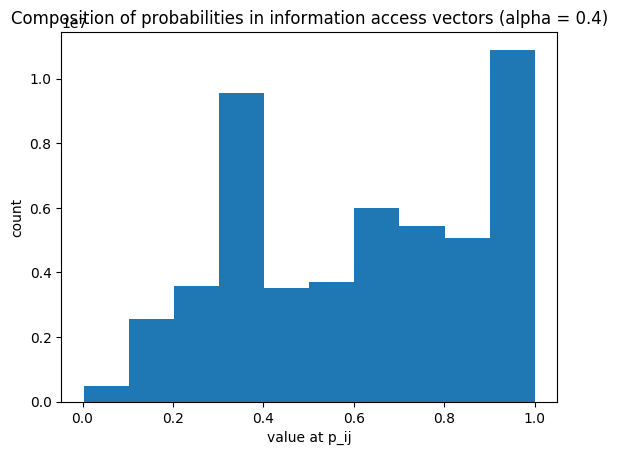} \\
$0.5$ & \includegraphics[align=c, height=1in]{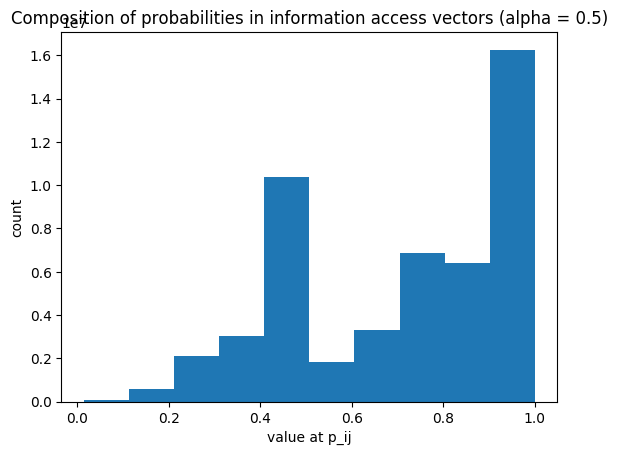} \\
$0.6$ & \includegraphics[align=c, height=1in]{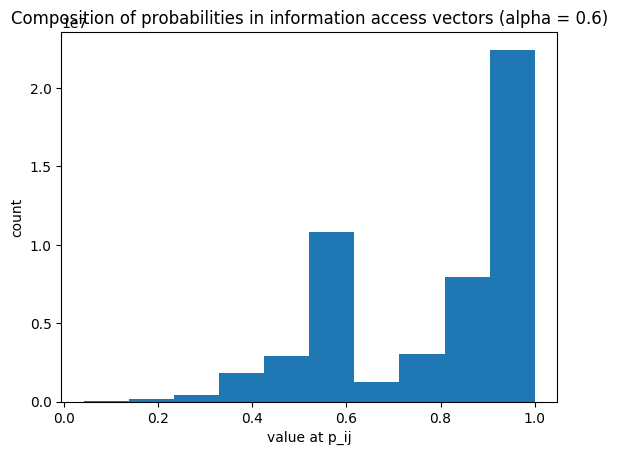} \\
$0.7$ & \includegraphics[align=c, height=1in]{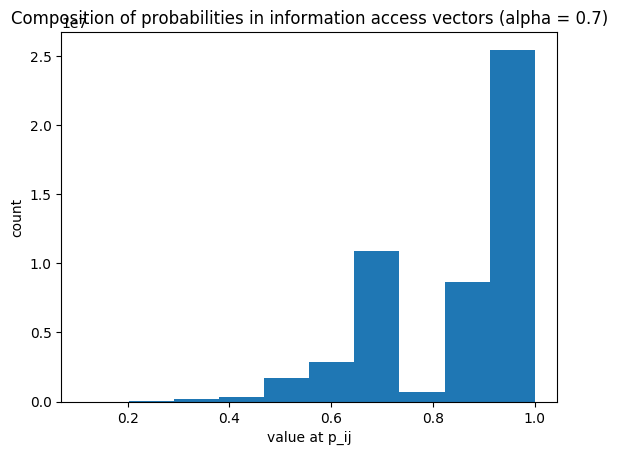} 
\end{tabular}
&
\begin{tabular}{c|c}
$\alpha$ & Co-sponsorship\\
$0.01$ & \includegraphics[align=c,height=1in]{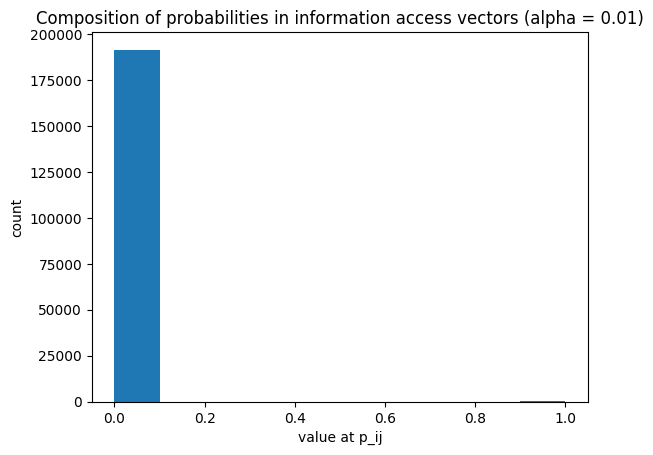}\\
$0.02$ & \includegraphics[align=c,height=1in]{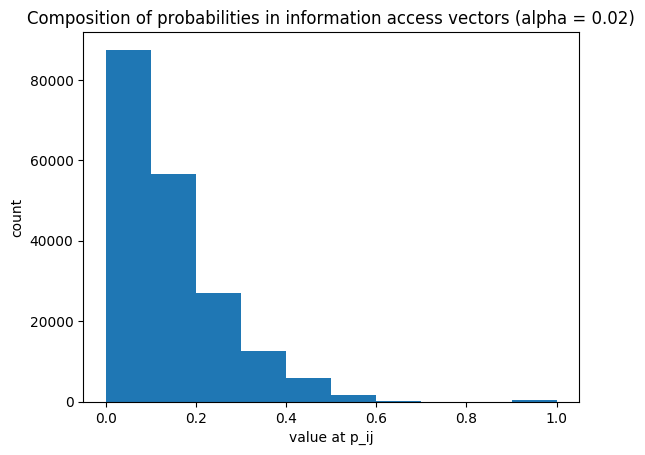}\\
$0.03$ & \includegraphics[align=c,height=1in]{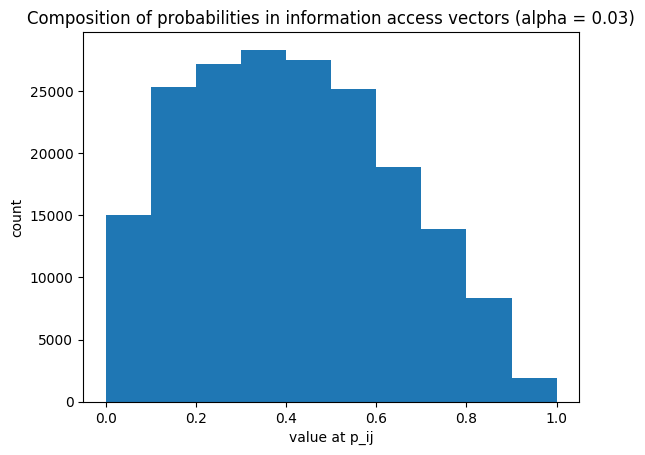}\\
$0.04$ & \includegraphics[align=c,height=1in]{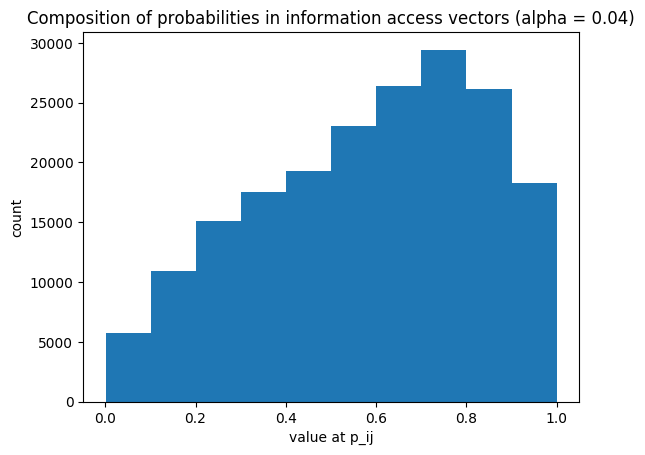}\\
$0.05$ & \includegraphics[align=c,height=1in]{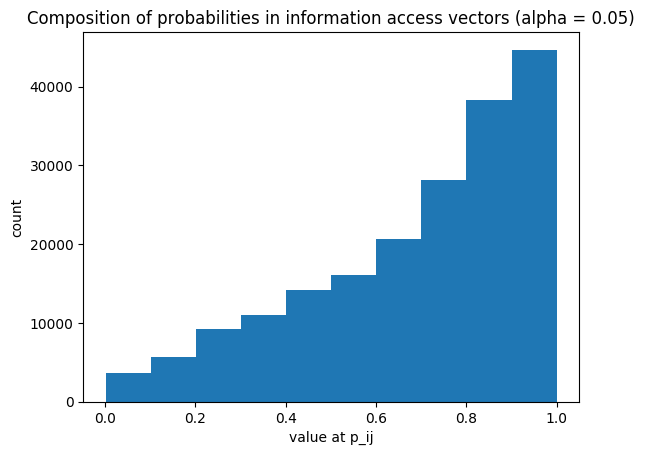}\\
$0.06$ & \includegraphics[align=c,height=1in]{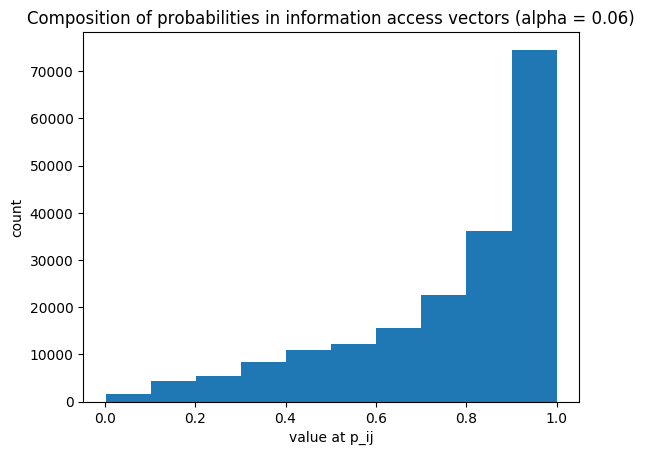}\\
$0.07$ & \includegraphics[align=c,height=1in]{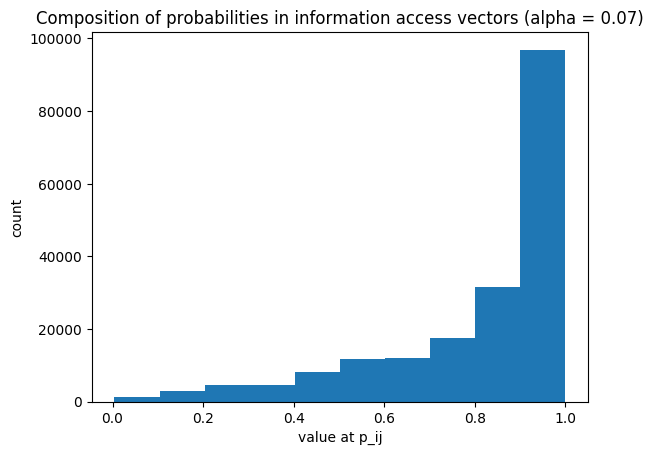}
\end{tabular}

\end{tabular}

\caption{Histograms showing the prevalence of $p_{ij}$ values within the information access signatures for the DBLP, Twitch, and Co-sponsorship data based on $\alpha$ value.}
\label{fig:histograms}
\end{center}
\end{figure*}

\newpage
\subsubsection{Clustering consistency across $\alpha$ ranges}
\label{apx:clustering_consistency}

\begin{figure}[htb]
\includegraphics[width=.3\columnwidth]{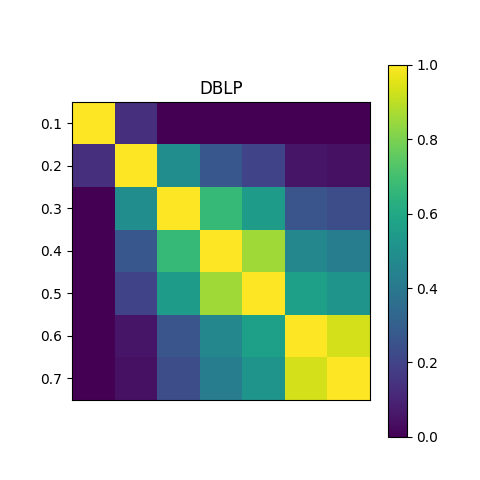}
\includegraphics[width=.3\columnwidth]{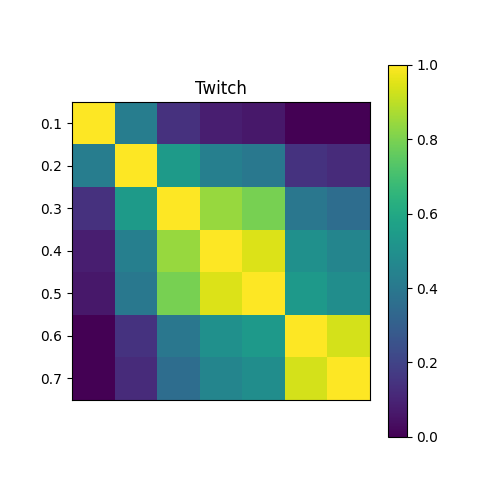}
\includegraphics[width=.315\columnwidth]{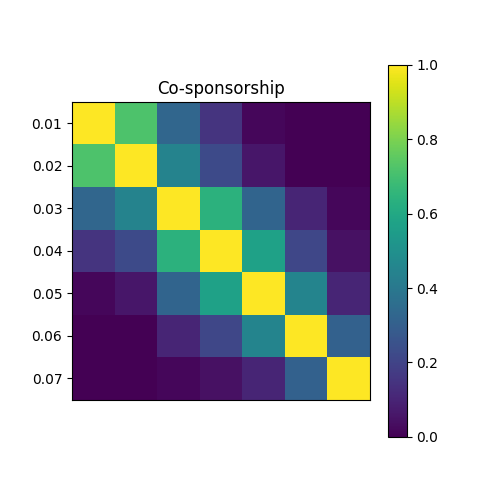}
\caption{A visualization of the consistency of clusterings across the ranges of $\alpha$ investigated for each dataset. Each visualization shows the Adjusted Rand Index (ARI). We round ARIs up to 0 when negative to ensure consistency across colorscales for the three datasets.\label{fig:consistency}}
\end{figure}

As an additional validation step, here we investigate the extent to which clusterings generated by the different $\alpha$ values are consistent across the $\alpha$ values investigated. We do not expect consistent clusterings whenever the probability vectors are significantly different from one another, but expect consistency between clusterings with similar $\alpha$ values. 
Figure~\ref{fig:consistency} shows the adjusted rand index values computed between the clustering assignments for all $\alpha$ values. Notice how for both the Twitch and DBLP datasets, there exists a clear range of $\alpha$ ($[0.2, 0.5]$ and $[0.3, 0.5]$ respectively) where the clusterings largely agree with one another. The situation is not as clear for the co-sponsorship dataset, but there, we can still see that similar $\alpha$ values tend to generate similar clustering assignments. Importantly, given the gradations along the diagonal it seems likely from this analysis that we are not missing important clusterings due to the choice of granularity of $\alpha$ values considered.

\newpage

\subsection{Choosing $k$}\label{apx:choose-k}
\begin{figure}[h!]
\begin{tabular}{ccc}
\begin{tabular}{r|c}
\multicolumn{2}{c}{\textbf{DBLP}}\\
$\alpha$ & Elbow Method\\
$0.1$ & \includegraphics[align=c,height=1.0in]{figs/dblp/elbow_method_plots/dblp_elbow_i1_10000.png}\\
$0.2$ & \includegraphics[align=c,height=1.0in]{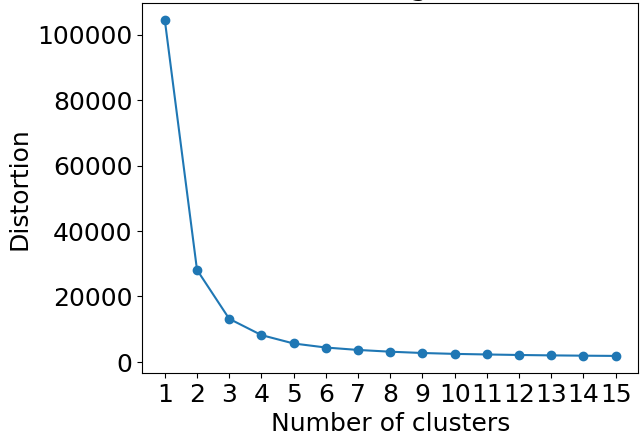}\\
$0.3$ & \includegraphics[align=c,height=1.0in]{figs/dblp/elbow_method_plots/dblp_elbow_i3_10000.png}\\
$0.4$ & \includegraphics[align=c,height=1.0in]{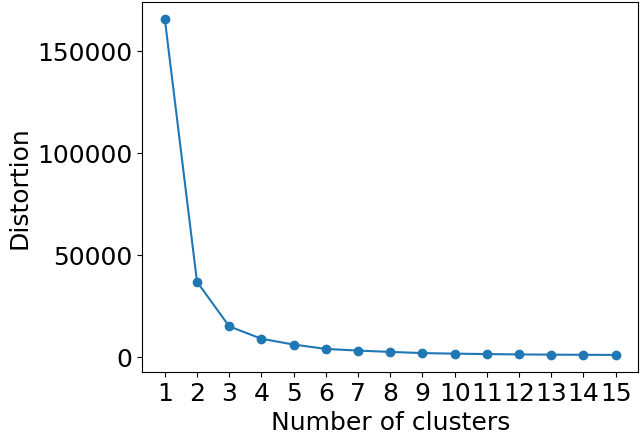}\\
$0.5$ & \includegraphics[align=c,height=1.0in]{figs/dblp/elbow_method_plots/dblp_elbow_i5_10000.png}\\
$0.6$ & \includegraphics[align=c,height=1.0in]{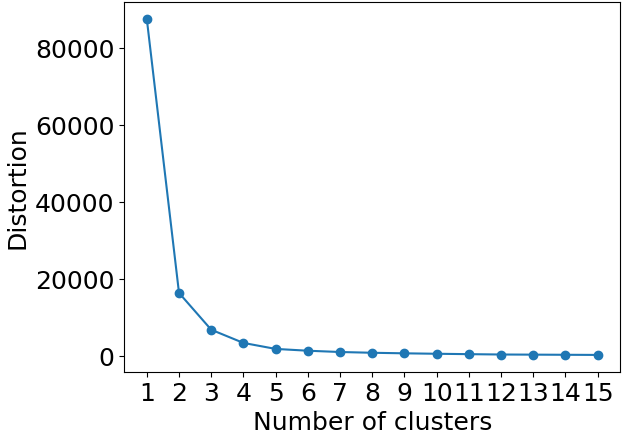}\\
$0.7$ & \includegraphics[align=c,height=1.0in]{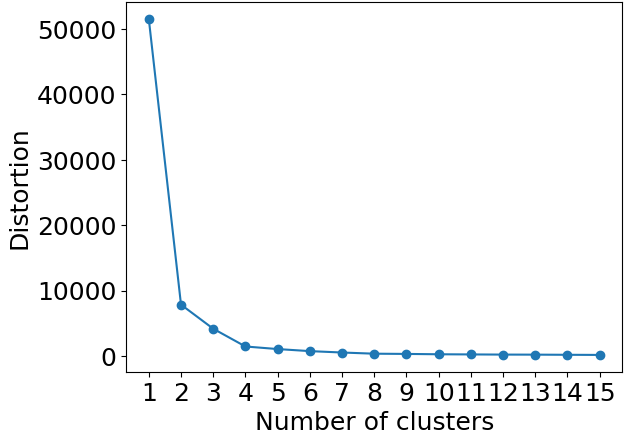}
\end{tabular}
&
\begin{tabular}{r|c}
\multicolumn{2}{c}{\textbf{Twitch}}\\
$\alpha$ & Elbow Method\\
$0.1$ & \includegraphics[align=c,height=1.0in]{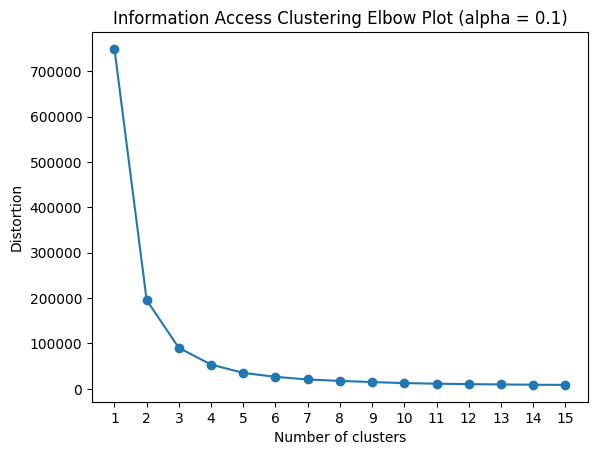}\\
$0.2$ & \includegraphics[align=c,height=1.0in]{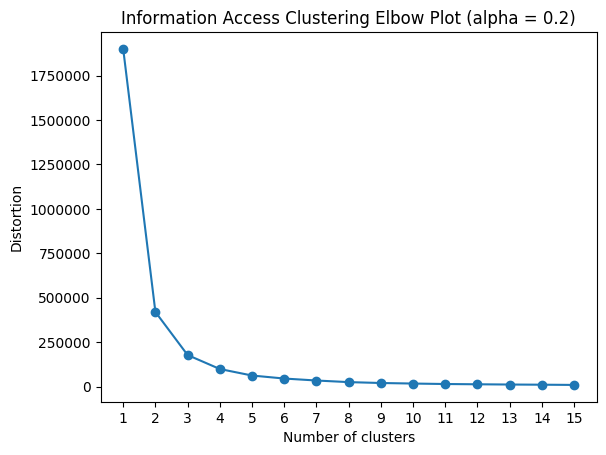}\\
$0.3$ & \includegraphics[align=c,height=1.0in]{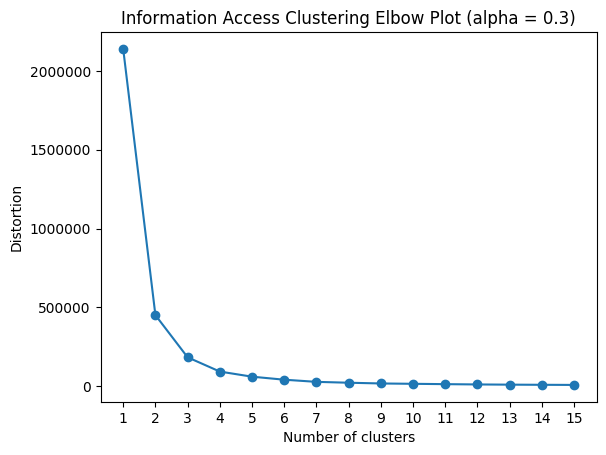}\\
$0.4$ & \includegraphics[align=c,height=1.0in]{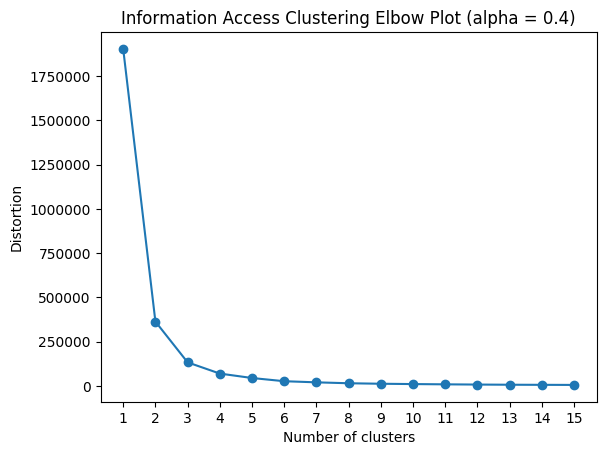}\\
$0.5$ & \includegraphics[align=c,height=1.0in]{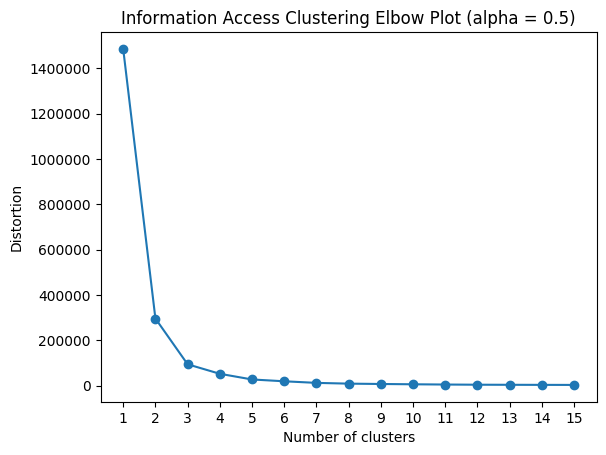}\\
$0.6$ & \includegraphics[align=c,height=1.0in]{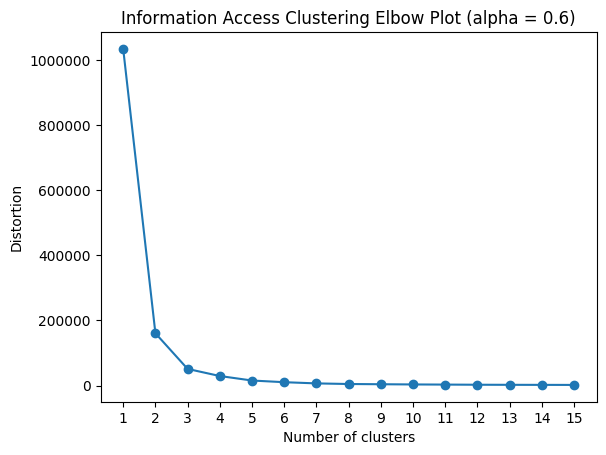}\\
$0.7$ & \includegraphics[align=c,height=1.0in]{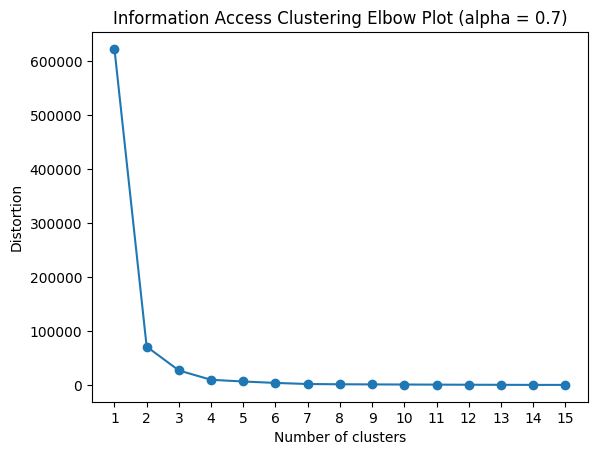}\\
\end{tabular}
&
\begin{tabular}{r|c}
\multicolumn{2}{c}{\textbf{Co-sponsorship}}\\
$\alpha$ & Elbow Method\\
$0.01$ & \includegraphics[align=c,height=1.0in]{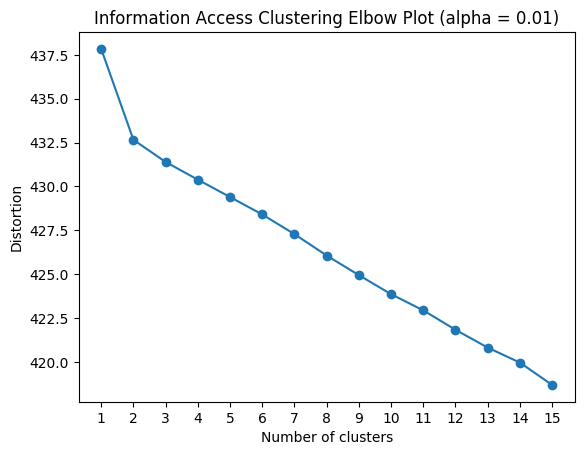}\\
$0.02$ & \includegraphics[align=c,height=1.0in]{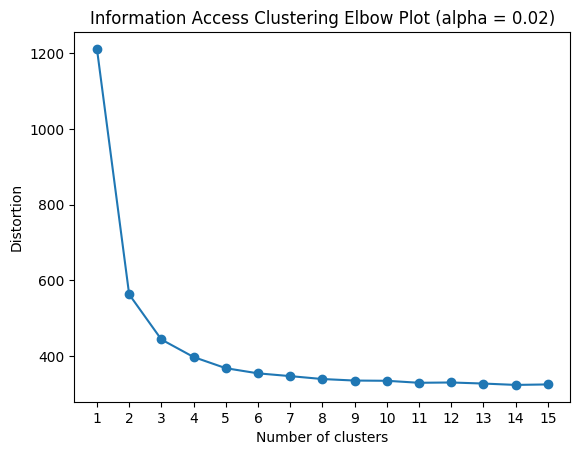}\\
$0.03$ & \includegraphics[align=c,height=1.0in]{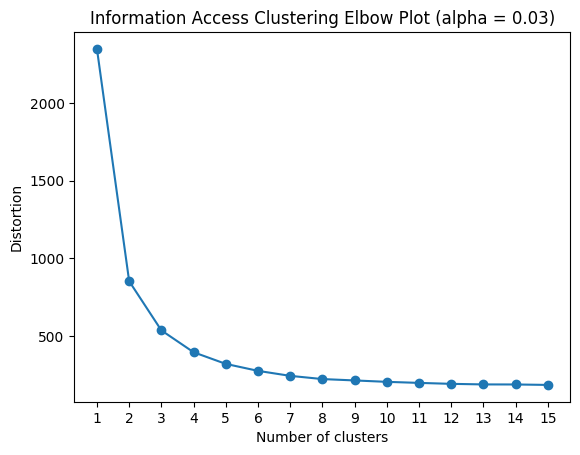}\\
$0.04$ & \includegraphics[align=c,height=1.0in]{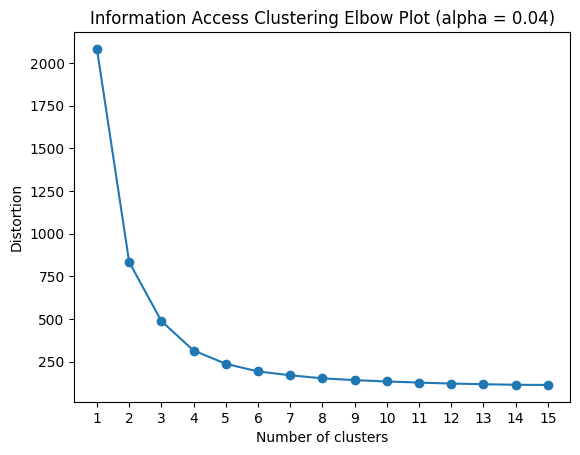}\\
$0.05$ & \includegraphics[align=c,height=1.0in]{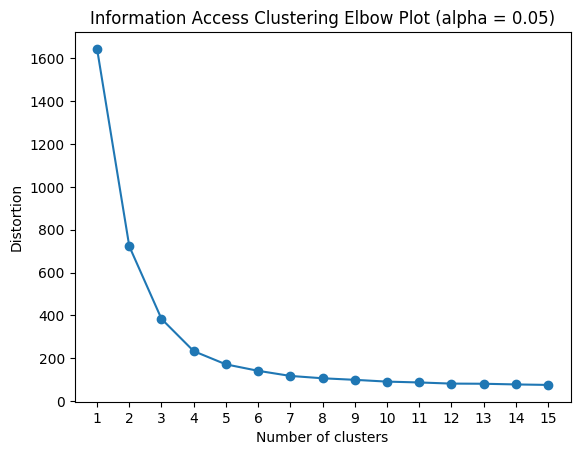}\\
$0.06$ & \includegraphics[align=c,height=1.0in]{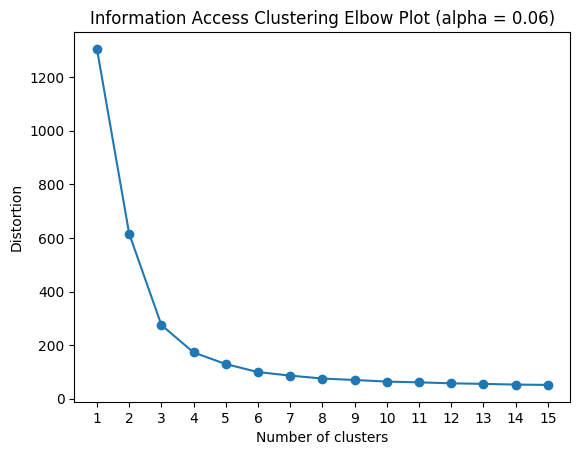}\\
$0.07$ & \includegraphics[align=c,height=1.0in]{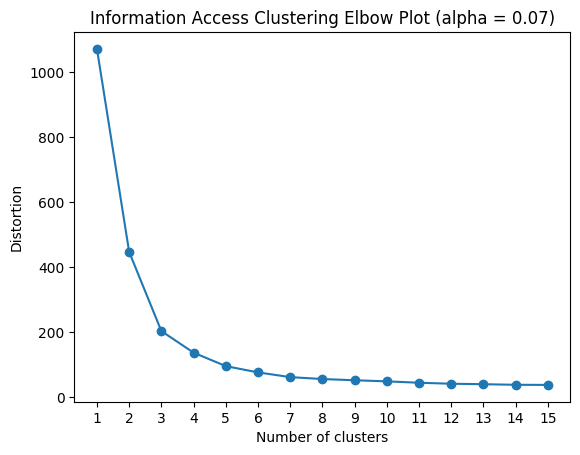}\\
\end{tabular}
\end{tabular}
\caption{Elbow method plots for information access clustering on the DBLP, Twitch, and Co-sponsorship datasets.}
\label{fig:elbow}
\end{figure}

\newpage

\begin{table}[h!]
\begin{center}
\begin{tabular}{c||ccccccccc}
\multicolumn{9}{c}{\textbf{DBLP}}\\
& \multicolumn{8}{c}{$k$} \\
$\alpha$ & 2 & 3 & 4 & 5 & 6 & 7 & 8 & 9 & 10\\
\hline
0.1 & \textbf{0.279} & 0.140 & 0.072 & 0.053 & 0.035 & 0.033 & 0.023 & 0.024 & 0.016 \\
0.2 & \textbf{0.598} & 0.551 & 0.503 & 0.479 & 0.449 & 0.411 & 0.389 & 0.374 & 0.355\\
0.3 & \textbf{0.658} & 0.605 & 0.578 & 0.558 & 0.551 & 0.532 & 0.517 & 0.501 & 0.485\\
0.4 & \textbf{0.714} & 0.692 & 0.651 & 0.646 & 0.621 & 0.605 & 0.596 & 0.591 & 0.581\\
0.5 & \textbf{0.753} & 0.740 & 0.717 & 0.723 & 0.676 & 0.657 & 0.644 & 0.640 & 0.631\\
0.6 & \textbf{0.828} & 0.787 & 0.796 & 0.783 & 0.778 & 0.772 & 0.728 & 0.726 & 0.721\\
0.7 & \textbf{0.876} & 0.836 & 0.847 & 0.823 & 0.829 & 0.820 & 0.821 & 0.821 & 0.774\\
\end{tabular}
~\\~\\~\\
\begin{tabular}{c||ccccccccc}
\multicolumn{9}{c}{\textbf{Twitch}}\\
& \multicolumn{8}{c}{$k$} \\
$\alpha$ & 2 & 3 & 4 & 5 & 6 & 7 & 8 & 9 & 10\\
\hline
0.1 & \textbf{0.640} & 0.582 & 0.538 & 0.510 & 0.484 & 0.462 & 0.432 & 0.409 & 0.404\\
0.2 & \textbf{0.652} & 0.620 & 0.600 & 0.594 & 0.581 & 0.571 & 0.569 & 0.564 & 0.544\\
0.3 & \textbf{0.682} & 0.667 & 0.671 & 0.665 & 0.668 & 0.665 & 0.656 & 0.641 & 0.620\\
0.4 & 0.732 & 0.730 & 0.729 & 0.733 & \textbf{0.734} & 0.728 & 0.721 & 0.720 & 0.711\\
0.5 & 0.753 & 0.764 & 0.763 & 0.767 & 0.765 & \textbf{0.769} & 0.759 & 0.752 & 0.751\\
0.6 & 0.822 & 0.823 & \textbf{0.828} & 0.815 & 0.817 & 0.816 & 0.805 & 0.793 & 0.802\\
0.7 & 0.872 & 0.864 & \textbf{0.873} & 0.852 & 0.860 & 0.850 & 0.851 & 0.830 & 0.811
\end{tabular}
~\\~\\~\\
\begin{tabular}{c||ccccccccc}
\multicolumn{9}{c}{\textbf{Co-sponsorship}}\\
& \multicolumn{8}{c}{$k$} \\
$\alpha$ & 2 & 3 & 4 & 5 & 6 & 7 & 8 & 9 & 10\\
\hline
0.01 & \textbf{0.009} & 0.004 & 0.002 & 0.0007 & -0.0003 & 0.0003 & 0.00003 & 0.0001 & 0.00002\\
0.02 & \textbf{0.411} & 0.268 & 0.188 & 0.153 & 0.125 & 0.096 & 0.081 & 0.074 & 0.061\\
0.03 & \textbf{0.538} & 0.446 & 0.397 & 0.377 & 0.308 & 0.290 & 0.260 & 0.254 & 0.216\\
0.04 & \textbf{0.584} & 0.543 & 0.498 & 0.429 & 0.397 & 0.388 & 0.335 & 0.272 & 0.266\\
0.05 & \textbf{0.658} & 0.613 & 0.550 & 0.497 & 0.490 & 0.401 & 0.381 & 0.320 & 0.317\\
0.06 & \textbf{0.738} & 0.664 & 0.584 & 0.554 & 0.499 & 0.508 & 0.417 & 0.396 & 0.377\\
0.07 & \textbf{0.899} & 0.721 & 0.600 & 0.588 & 0.500 & 0.500 & 0.475 & 0.378 & 0.437
\end{tabular}
\caption{Silhouette values for varying $\alpha$ and $k$ for the DBLP, Twitch, Co-sponsorship, and Google Scholar datasets for information access clustering.  The largest silhouette value for each $\alpha$ is shown in bold.}
\label{fig:silhouette}
\end{center}
\end{table}

\newpage

\section{Clusters and external information access measures}
\label{apx:prob_density}

\begin{figure}[h!]
\begin{center}
\begin{tabular}{c|ccc}
& \multicolumn{3}{c}{\textbf{DBLP}} \\
$\alpha$ & ~~~~~~~Citation Count & ~~~~~~~~~~PhD Rank & ~~~~~~Job Rank\\
$0.1$ & \multicolumn{3}{c}{\includegraphics[align=c, height=1.1in]{figs/dblp/prob_densities/alpha0.1.png}} \\
$0.2$ & \multicolumn{3}{c}{\includegraphics[align=c, height=1.1in]{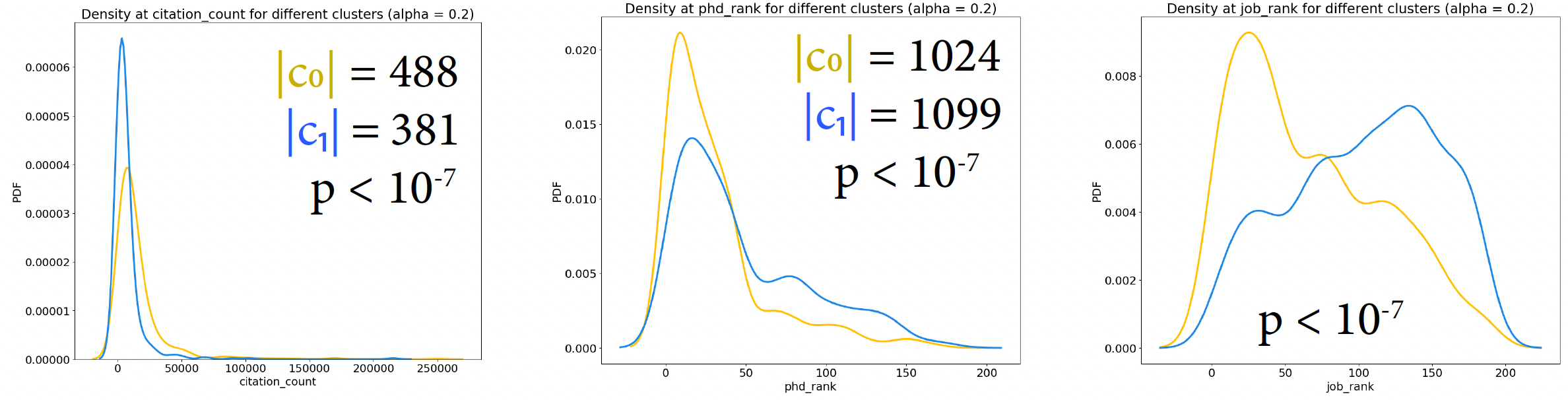}} \\
$0.3$ & \multicolumn{3}{c}{\includegraphics[align=c, height=1.1in]{figs/dblp/prob_densities/alpha0.3.png}} \\
$0.4$ & \multicolumn{3}{c}{\includegraphics[align=c, height=1.1in]{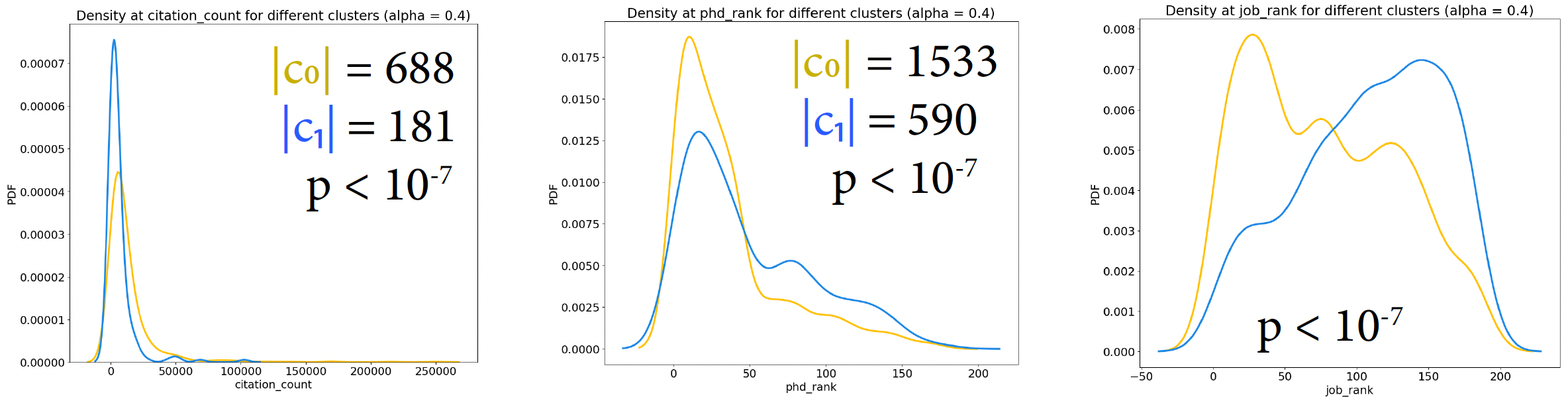}} \\
$0.5$ & \multicolumn{3}{c}{\includegraphics[align=c, height=1.1in]{figs/dblp/prob_densities/alpha0.5.png}} \\
$0.6$ & \multicolumn{3}{c}{\includegraphics[align=c, height=1.1in]{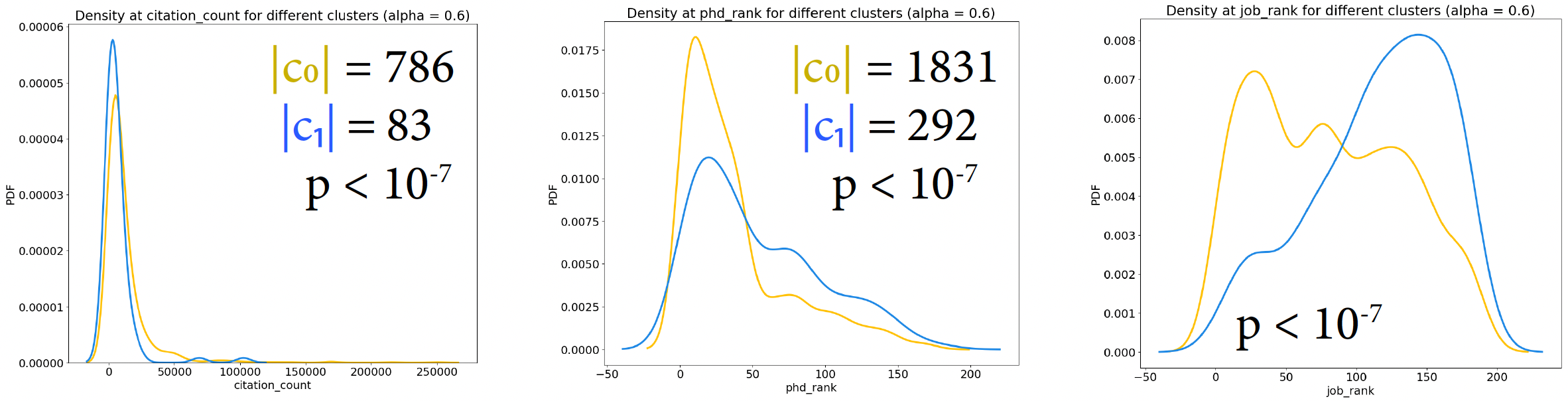}} 
\end{tabular}
\end{center}
\caption{Results above are shown for the DBLP dataset.  Clusterings are given for $\alpha$ values as shown on the left.  The probability density function (PDF) per cluster is shown with respect to the indicated external information access measure given on the $x$-axis.
  Note that the cluster counts for the Job Rank column are the same as for the PhD Rank column; the annotations are omitted in the figure to avoid obscuring the distribution curves.}
\label{fig:dblp}
\end{figure}

\begin{figure*}[htbp]

\begin{center}
\begin{tabular}{cc}

\begin{tabular}{c|cc}
& \multicolumn{2}{c}{\textbf{Twitch}} \\
$\alpha$ & ~~~~~~~~~~~~~Partner & ~~~~~~~~~~log(Views) \\
$0.1$ & \multicolumn{2}{c}{\includegraphics[align=c, height=1.1in]{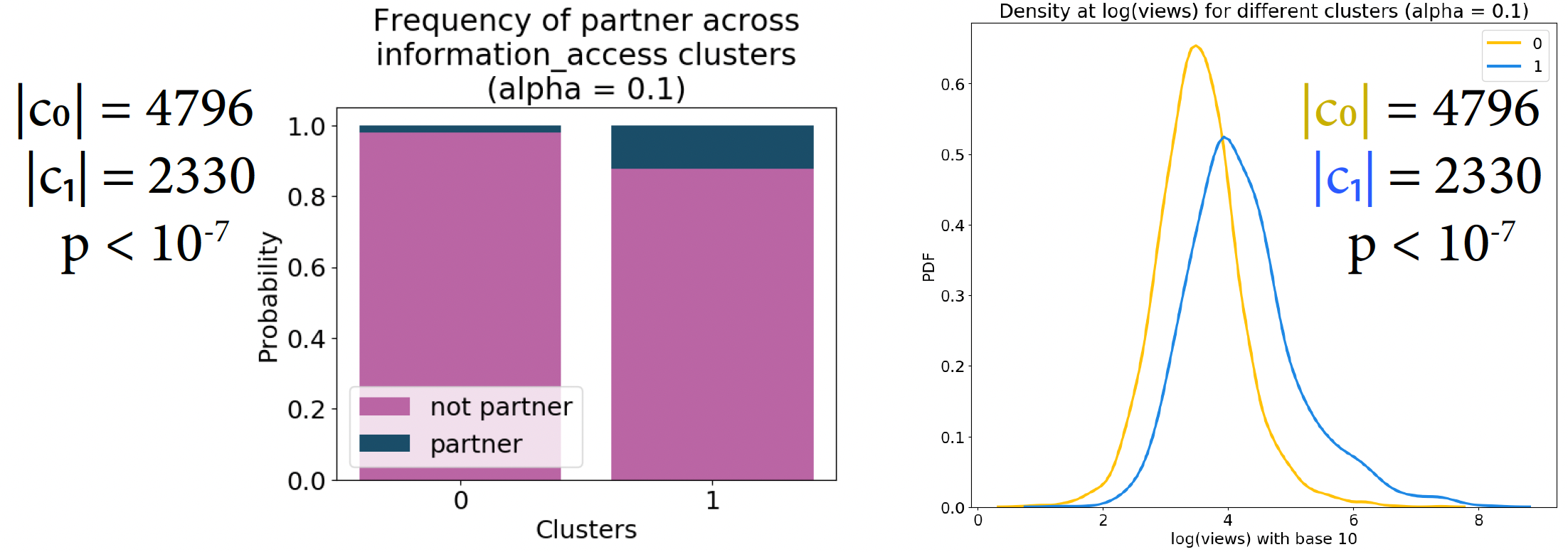}} \\
$0.2$ & \multicolumn{2}{c}{\includegraphics[align=c, height=1.1in]{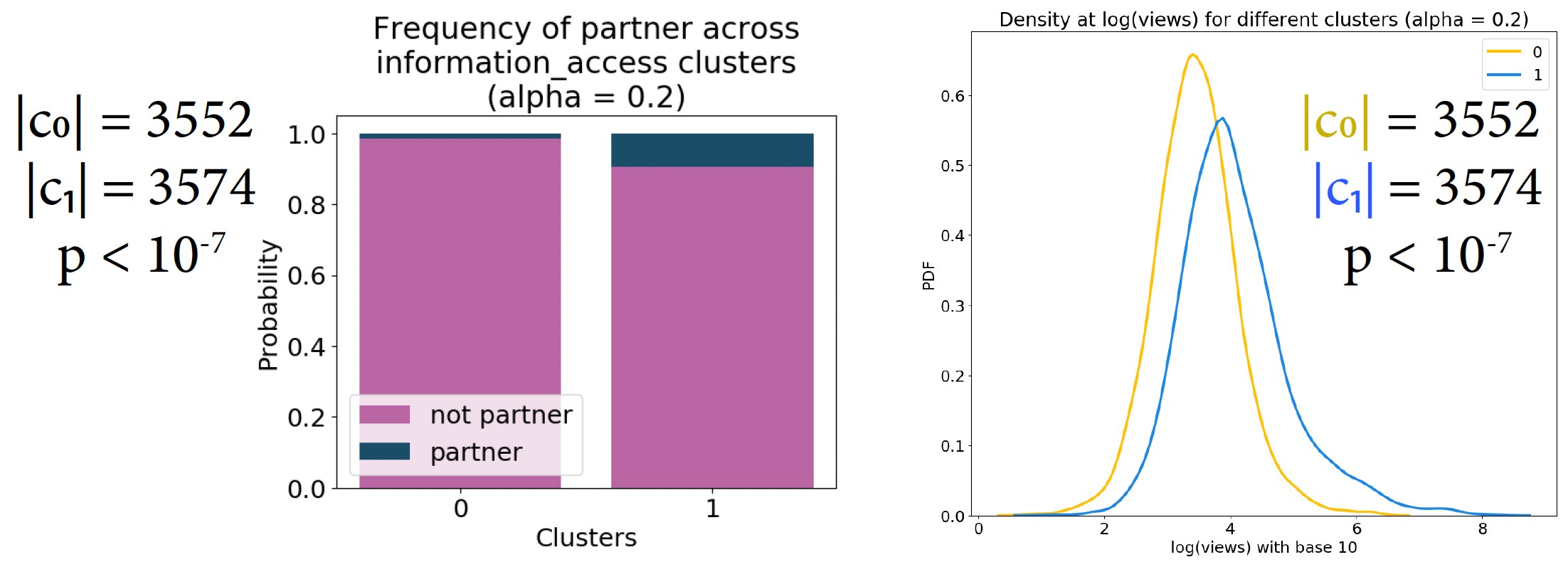}} \\
$0.3$ & \multicolumn{2}{c}{\includegraphics[align=c, height=1.1in]{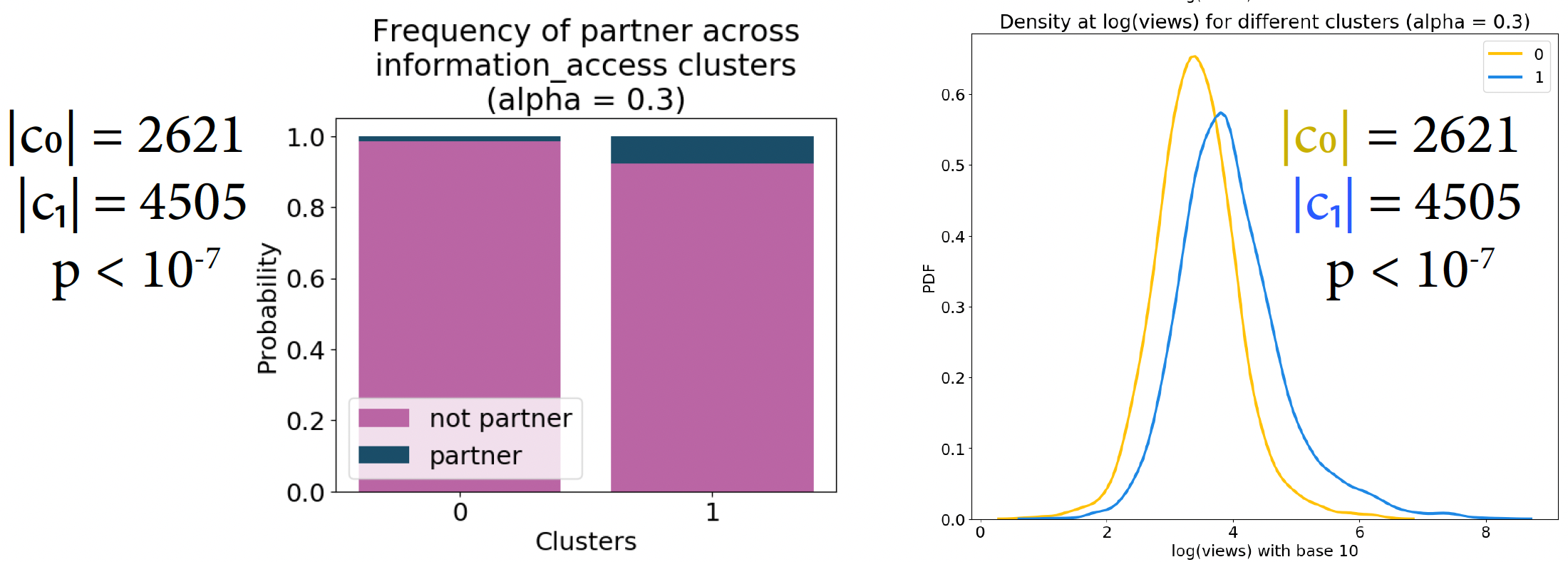}} \\
$0.4$ & \multicolumn{2}{c}{\includegraphics[align=c, height=1.1in]{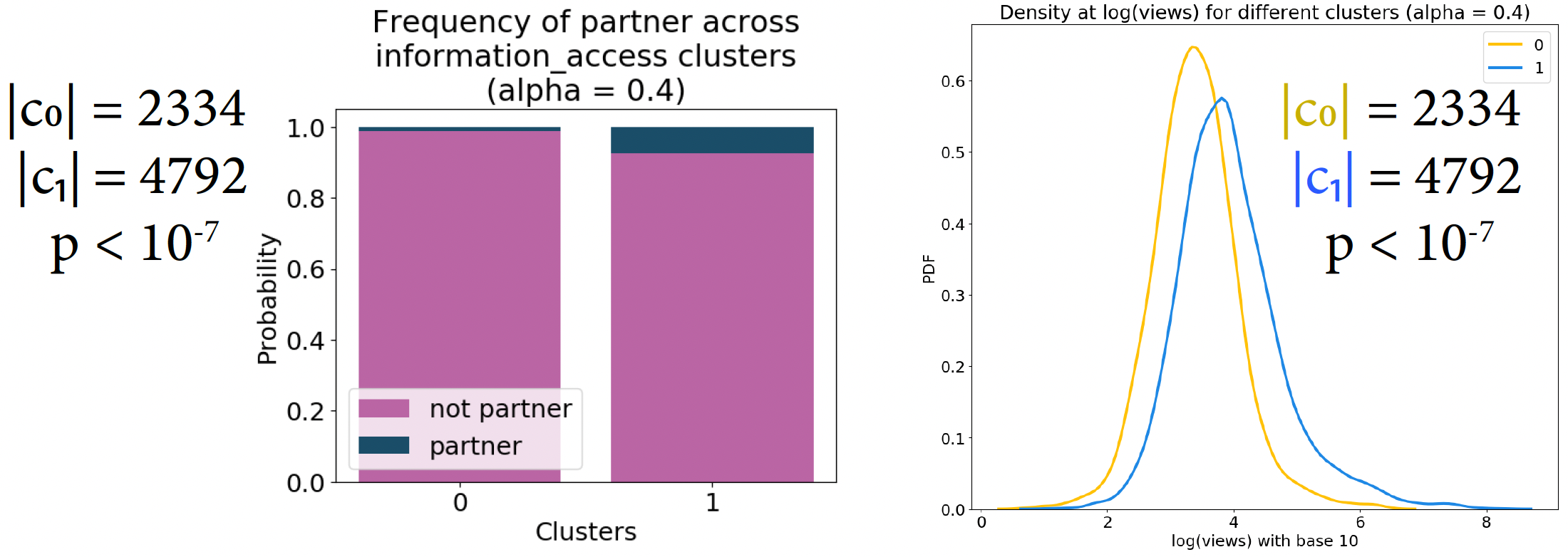}} \\
$0.5$ & \multicolumn{2}{c}{\includegraphics[align=c, height=1.1in]{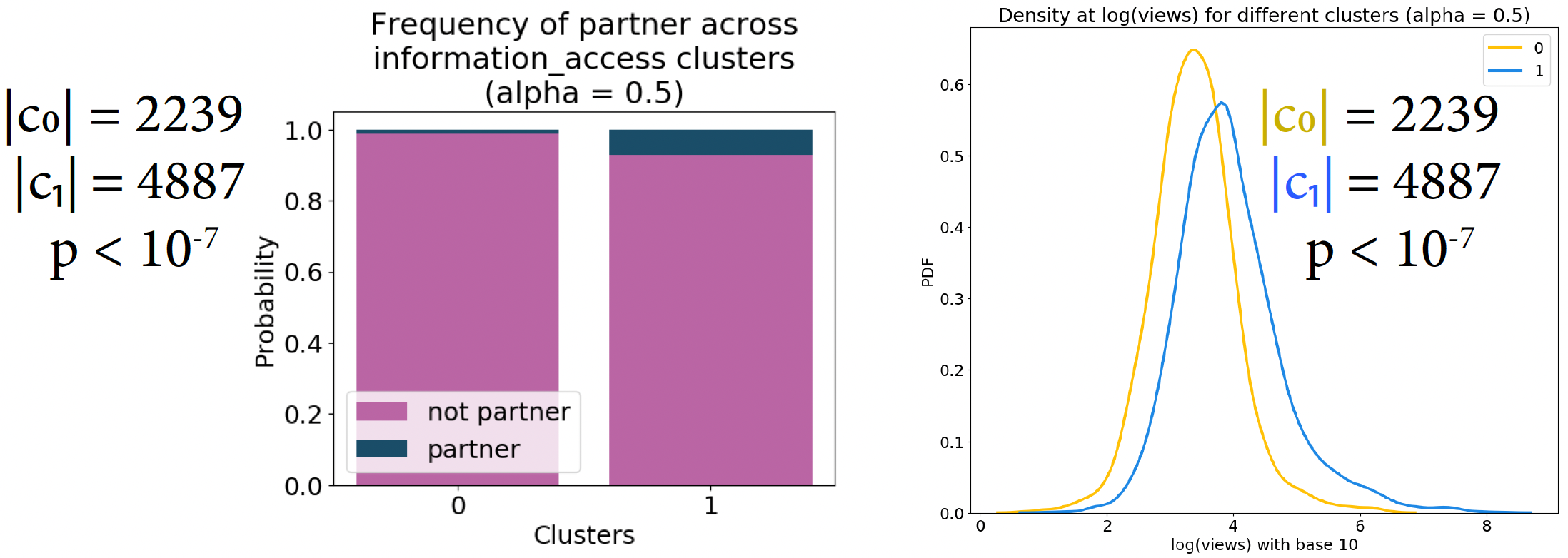}} \\
$0.6$ & \multicolumn{2}{c}{\includegraphics[align=c, height=1.1in]{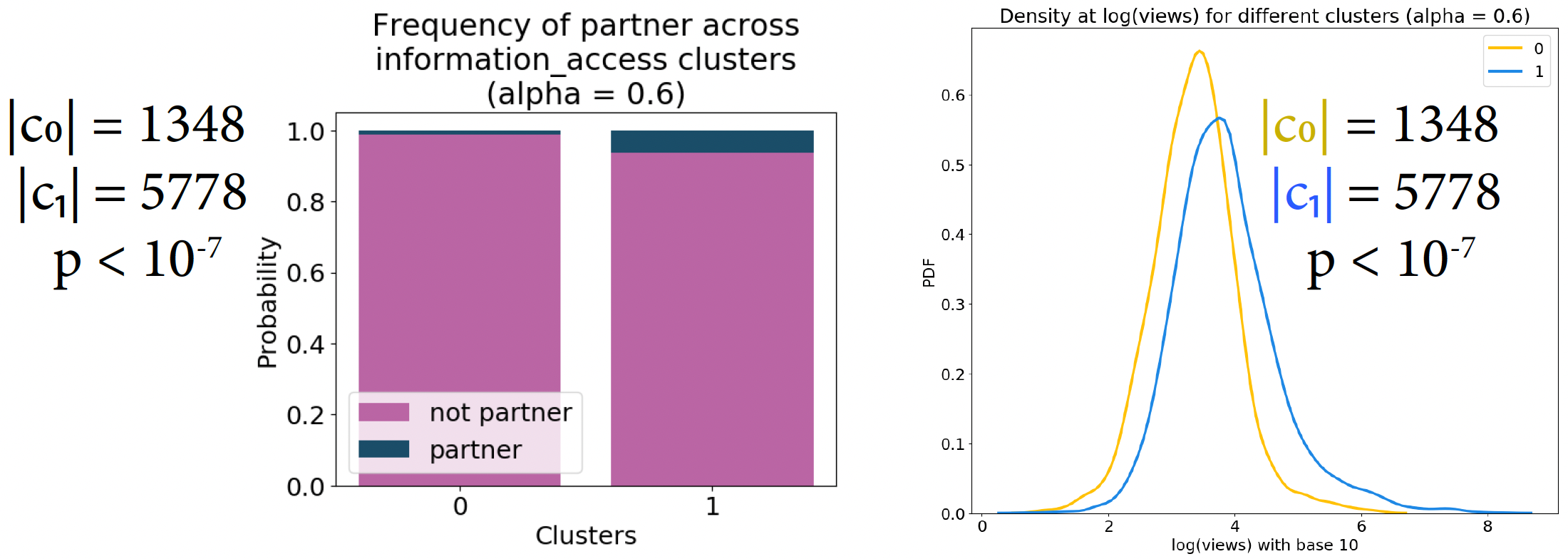}} 
\end{tabular}

&~~~~~

\begin{tabular}{c|c}
& \textbf{Co-sponsorship} \\
$\alpha$ & Legislative Effectiveness \\
$0.01$ & \includegraphics[align=c, height=1.1in]{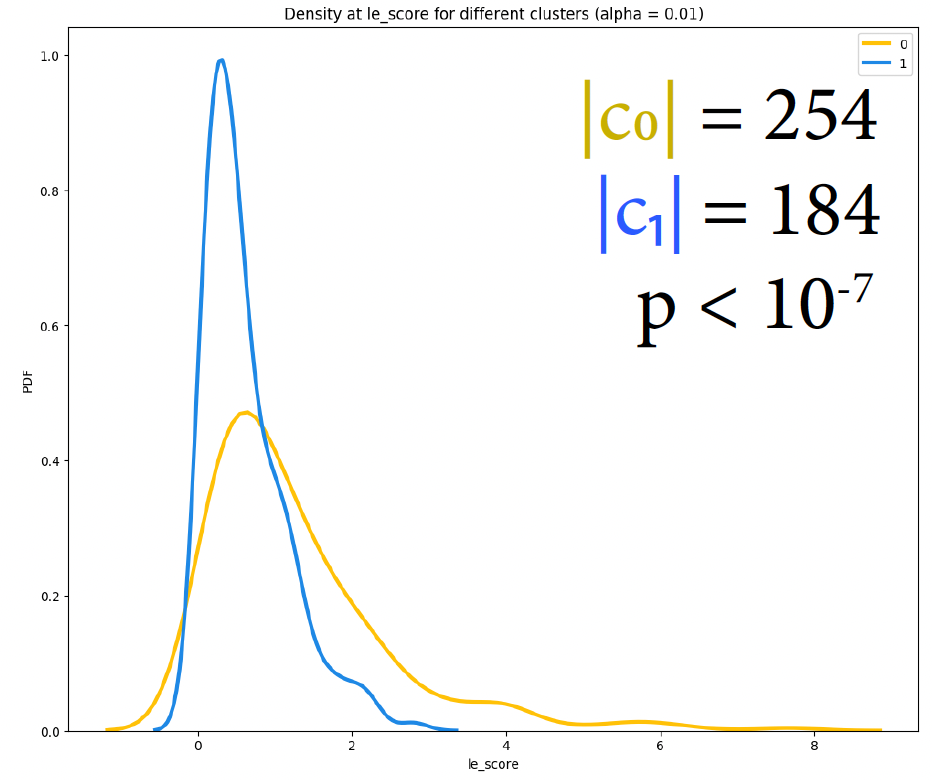} \\
$0.02$ & \includegraphics[align=c, height=1.1in]{figs/cosponsorship/prob_densities/alpha0.01.png} \\
$0.03$ & \includegraphics[align=c, height=1.1in]{figs/cosponsorship/prob_densities/alpha0.01.png} \\
$0.04$ & \includegraphics[align=c, height=1.1in]{figs/cosponsorship/prob_densities/alpha0.01.png} \\
$0.05$ & \includegraphics[align=c, height=1.1in]{figs/cosponsorship/prob_densities/alpha0.01.png} \\
$0.06$ & \includegraphics[align=c, height=1.1in]{figs/cosponsorship/prob_densities/alpha0.01.png} 
\end{tabular}
\end{tabular}

\end{center}
\caption{Results above are shown for the Twitch and Co-sponsorship datasets.  Clusterings are given for $\alpha$ values as shown on the left.  The probability density function (PDF) or composition per cluster is shown with respect to the indicated external information access measure.}
\label{fig:twitch_cospons}
\end{figure*}

\newpage

\section{Comparison between different clustering methods}
\label{sec:apx-comparisons}

\begin{table}[h!]
\begin{center}
\textbf{Spectral Clustering}
\begin{tabular}{ccc}
\begin{tabular}{c|c}
$\alpha$ & DBLP \\
\hline
$0.1$  & \smallari \\ 
$0.2$  & \smallari \\ 
$0.3$  & \smallari \\ 
$0.4$  & \smallari \\ 
$0.5$  & \smallari \\ 
$0.6$  & 0.02 \\ 
$0.7$  & 0.02 \\ 
\end{tabular}

&
\begin{tabular}{c|c}
$\alpha$ & Twitch \\
\hline
0.1 & \smallari \\ 
0.2 &\smallari \\ 
0.3 &\smallari \\ 
0.4 & \smallari \\ 
0.5 & \smallari \\ 
0.6 & \smallari \\ 
0.7 & \smallari \\ 
\end{tabular}
&

\begin{tabular}{c|c}
$\alpha$ & Co-spons. \\
\hline
0.01 & \smallari \\ 
0.02 &\smallari \\ 
0.03 & \smallari \\ 
0.04 & \smallari \\ 
0.05 & \smallari \\ 
0.06 & \smallari \\ 
0.07 & \smallari \\ 
\end{tabular}
\end{tabular}

~\\
\textbf{Fluid Communities}
\begin{tabular}{ccc}
\begin{tabular}{c|c}
$\alpha$ & DBLP \\
\hline
$0.1$  & \smallari \\ 
$0.2$  & \smallari \\ 
$0.3$  & \smallari \\ 
$0.4$  &\smallari \\ 
$0.5$  & \smallari \\ 
$0.6$  & \smallari \\ 
$0.7$  & \smallari \\ 
\end{tabular}

&
\begin{tabular}{c|c}
$\alpha$ & Twitch \\
\hline
0.1 & \smallari \\ 
0.2 & 0.01 \\ 
0.3 & \smallari \\ 
0.4 & \smallari \\ 
0.5 & \smallari \\ 
0.6 & \smallari \\ 
0.7 & \smallari \\ 
\end{tabular}
&

\begin{tabular}{c|c}
$\alpha$ & Co-spons. \\
\hline
0.01 & 0.65 \\ 
0.02 & 0.50 \\ 
0.03 & 0.28 \\ 
0.04 & 0.17 \\ 
0.05 & 0.07 \\ 
0.06 & 0.02 \\ 
0.07 & \smallari \\ 
\end{tabular}
\end{tabular}

~\\
\textbf{Louvain}
\begin{tabular}{ccc}
\begin{tabular}{c|c}
$\alpha$ & DBLP \\
\hline
$0.1$  &\smallari \\ 
$0.2$  &\smallari \\ 
$0.3$  &\smallari \\ 
$0.4$  &\smallari \\ 
$0.5$  & 0.01 \\ 
$0.6$  & 0.03 \\ 
$0.7$  & 0.01 \\ 
\end{tabular}

&
\begin{tabular}{c|c}
$\alpha$ & Twitch \\
\hline
0.1 & \smallari \\ 
0.2 &\smallari \\ 
0.3 &\smallari \\ 
0.4 & \smallari \\ 
0.5 &\smallari \\ 
0.6 & \smallari \\ 
0.7 &\smallari \\ 
\end{tabular}
&

\begin{tabular}{c|c}
$\alpha$ & Co-spons. \\
\hline
0.01 & 0.85 \\
0.02 & 0.59 \\
0.03 & 0.27 \\
0.04 & 0.12 \\
0.05 & 0.02\\
0.06 & \smallari \\
0.07 & \smallari \\
\end{tabular}
\end{tabular}

~\\
\textbf{Role2Vec}
\begin{tabular}{ccc}
\begin{tabular}{c|c}
$\alpha$ & DBLP \\
\hline
$0.1$  & \smallari \\
$0.2$  & 0.37 \\
$0.3$  & 0.49\\
$0.4$  & 0.44 \\
$0.5$  & 0.38 \\
$0.6$  & 0.22 \\
$0.7$  & 0.20
\end{tabular}

&
\begin{tabular}{c|c}
$\alpha$ & Twitch \\
\hline
0.1 & 0.11 \\
0.2 & 0.15\\
0.3 & 0.13\\
0.4 & 0.12 \\
0.5 & 0.11\\
0.6 & 0.05\\
0.7 & 0.04 \\
\end{tabular}
&

\begin{tabular}{c|c}
$\alpha$ & Co-spons. \\
\hline
0.01 & 0.77 \\
0.02 & 0.55 \\
0.03 & 0.27 \\
0.04 & 0.13 \\
0.05 & 0.03 \\
0.06 & \smallari \\
0.07 & \smallari \\
\end{tabular}
\end{tabular}

~\\
\textbf{Core-Periphery}
\begin{tabular}{ccc}
\begin{tabular}{c|c}
$\alpha$ & DBLP \\
\hline
$0.1$  & 0.73 \\
$0.2$  & 0.19 \\
$0.3$  & \smallari \\
$0.4$  & \smallari \\
$0.5$  & \smallari \\
$0.6$  & \smallari \\
$0.7$  & \smallari
\end{tabular}

&
\begin{tabular}{c|c}
$\alpha$ & Twitch \\
\hline
0.1 & 0.54 \\
0.2 & 0.16 \\
0.3 & \smallari \\
0.4 &  \smallari \\
0.5 & \smallari \\
0.6 & \smallari \\
0.7 & \smallari \\
\end{tabular}
&

\begin{tabular}{c|c}
$\alpha$ & Co-spons. \\
\hline
0.01 & 0.12 \\
0.02 & 0.11 \\
0.03 & \smallari \\
0.04 & \smallari \\
0.05 & \smallari \\
0.06 & \smallari \\
0.07 & \smallari \\
\end{tabular}
\end{tabular}

\end{center}

\caption{For each dataset, $\alpha$-parameterized clustering, and $k$ clusters, the above table gives the adjusted rand index indicating the difference between the resulting information access and the indicated clustering method.  An adjusted rand index of $0$ indicates only random agreement between clusterings while a value of $1$ indicates an exact match. For legibility, adjusted rand index values less than $0.01$ are shown as $\sim 0$. Fluid communities method given values are the mean across $10$ random runs.}

\label{tab:comparison_aris}
\end{table}

\section{Large networks}

\begin{table}[h!]
\begin{center}
\begin{tabular}{c|cccccccccccccc}
\multicolumn{15}{c}{\textbf{Random sampling}}\\
$\alpha$ & \multicolumn{14}{c}{Number of sampled seeds}\\
& 5	&10	&15	&20	&25	&30	&35	&40	&45&	50&	55&	60&	65&	70\\
\hline
0.1&	0.00&	0.81	&0.88&	0.89	&0.87	&0.83&	0.87&	0.92&	0.93	&0.92	&0.92&	0.88	&0.93	&0.92\\
0.2	&0.97&	0.99&	0.99&	0.98&	0.98	&0.99&	0.99&	0.99	&0.98	&0.99&	1 &	0.99&	0.99&	0.99\\
0.3&	0.97	&0.99&	0.99&	0.99	&0.99	&0.99	&0.99&	1	& 1 &0.99	&1	&0.99	&0.99	&0.99\\
0.4	&0.99&	0.99&	1	&0.99	&1	&1	&1	&1 &	0.99	&1&	1&	1&	1&	1\\
0.5&	1&	1	&1&	0.99&	1&	1	&1&	1&	1&	1&	1&	1&	1&	1\\
0.6	&0.99	&1&	1&	1&	1	&1	&1&	1	&1	&0.99&	0.99&	1&	0.99&	0.99\\
0.7	&1	&1&	1&	1	&1	&1&	1	&1	&1&	1&	1	&1	&1&	1
\end{tabular}
~\\~\\~\\

\begin{tabular}{c|cccccccccccccc}
\multicolumn{15}{c}{\textbf{Selection in order of PageRank}}\\
$\alpha$ & \multicolumn{14}{c}{Number of sampled seeds}\\
\hline
0.1	& 0.89	& 0.90	& 0.93	& 0.92	& 0.93	& 0.92	& 0.93	& 0.95	& 0.94	& 0.94	& 0.95	& 0.95	& 0.95&	0.94\\
0.2&	0.99&	0.99&	1&	1&	1&	1	&1	&1&	0.99&	1	&1&	1&	1&	1\\
0.3&	0.99	&1&	1&	0.99&	1&	0.99	&1&	1	&1	&1	&1&	0.99	&1&	1\\
0.4	&1&	1&	1&	1	&1&	1&	1	&0.99&	1&	0.99&	1	&0.99&	0.99&	1\\
0.5&	1&	1&	1&	1&	1&	1&	1&	1&	1&	1&	1&	1&	1&	1\\
0.6&	0.99	&0.99	&1&	0.99&	1	&1	&0.99	&1	&1&	0.99&	0.99&	1&	1	&1\\
0.7&	1&	1&	1&	1&	1&	1&	1&	1&	1&	1&	1&	1&	1&	1
\end{tabular}
~\\~\\~\\

\begin{tabular}{c|cccccccccccccc}
\multicolumn{15}{c}{\textbf{Selection in order of degree centrality}}\\
$\alpha$ & \multicolumn{14}{c}{Number of sampled seeds}\\
\hline
0.1	&0.901	&0.91&	0.92&	0.92	&0.95	&0.96	&0.94	&0.95	&0.94	&0.95&	0.95	&0.95&	0.95&	0.95\\
0.2&	1&	0.99&	0.99	&1	&0.99&	1&	1&	1	&1	&1	&1	&0.99	&1	&1\\
0.3	&1	&0.99	&1	&0.99	&1&	1	&0.99&	1&	1&	1	&1	&0.99	&1	&1\\
0.4	&1	&1&	1&	1&	1&	0.99	&1	&1&	1	&1	&1&	1	&1&	1\\
0.5	&1	&1&	1&	1&	1	&1&	1&	1&	1&	1&	1&	1&	1&	1\\
0.6	&1&	1&	1	&0.99&	1&	1	&0.99	&0.99	&0.99	&1	&1	&1	&1	&0.99\\
0.7	&1	&1&	1	&1	&1&	1&	1	&1&	1&	1	&1&	1&	1	&1
\end{tabular}
~\\~\\~\\

\begin{tabular}{c|cccccccccccccc}
\multicolumn{15}{c}{\textbf{Selection in order of betweenness centrality}}\\
$\alpha$ & \multicolumn{14}{c}{Number of sampled seeds}\\
\hline
0.1	&0.91	&0.91	&0.91	&0.91	&0.91&	0.92&	0.94	&0.92	&0.93	&0.92	&0.92	&0.93	&0.92&	0.92\\
0.2	&0.99	&0.99	&1&	1&	1	&1	&1	&0.99&	1&	1	&1&	1&	1&	1\\
0.3	&0.99	&0.99	&1	&0.99&	0.99&	1	&1	&1	&1	&1	&1&	1	&1&	1\\
0.4	&1&	1	&1	&1	&1	&1&	1	&1&	1&	1&	1&	1&	1	&1\\
0.5	&1	&1&	1&	1&	1&	1	&1	&1&	1	&1	&1&	1&	1&	1\\
0.6	&1	&1	&0.99&	1&	0.99&	1	&1	&0.99	&0.99	&0.99&	1	&0.99&	1	&0.99\\
0.7	&1	&1&	1&	1	&1	&1&	1	&1&	1&	1&	1&	1&	1	&1
\end{tabular}
\end{center}
\caption{Adjusted rand index values for the comparison between the information access clustering on the full representation for the DBLP network versus the information access clustering based on the shown selected number of seeds and seed selection strategies.  Adjusted rand index values are $1$ when two clusterings agree on all pairs of points.}
\label{fig:sampling_aris}
\end{table}

\newpage
\subsection{Google Scholar dataset experimental setup details}
\label{apx:gscholar}

\begin{figure}[htbp]
\begin{center}
\begin{tabular}{cccc}
$\alpha = 0.01$ & $\alpha = 0.03$ & $\alpha = 0.05$ & $\alpha = 0.4$ \\
 \includegraphics[align=c,width=1.5in]{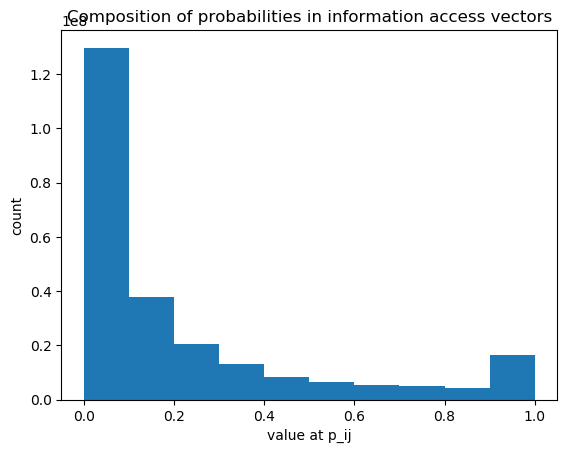}
 &
 \includegraphics[align=c,width=1.5in]{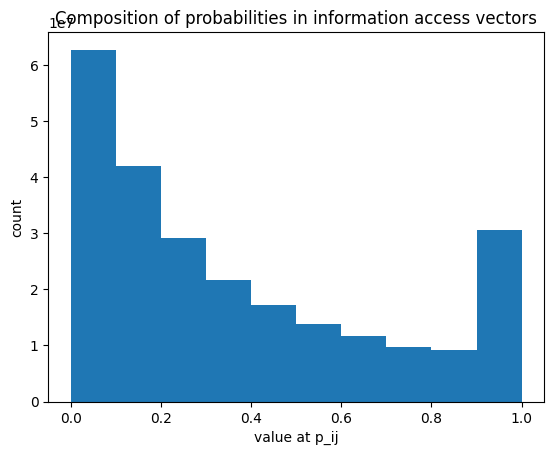}
&
 \includegraphics[align=c,width=1.5in]{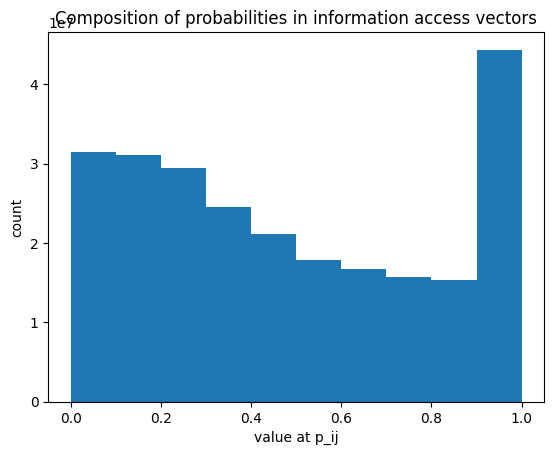}
&
 \includegraphics[align=c,width=1.5in]{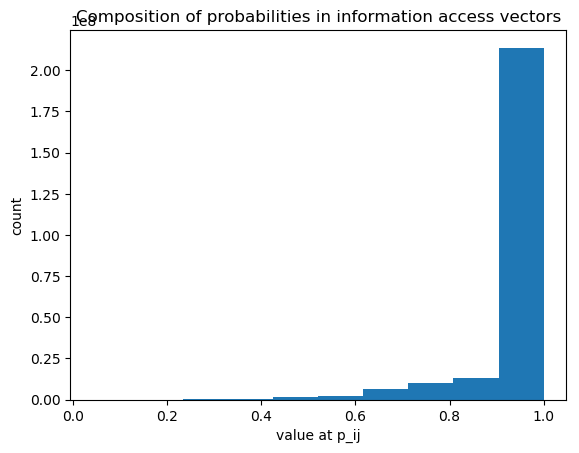}
\end{tabular}
\end{center}
\caption{Histograms showing the prevalence of $p_{ij}$ values within the information access signatures for the Google Scholar data based on $\alpha$ value.}
\label{fig:gscholar_pdist}
\end{figure}

\begin{figure*}[htbp]
\begin{center}
\begin{tabular}{cccc}
$\alpha = 0.01$ & $\alpha = 0.03$ & $\alpha = 0.05$ & $\alpha = 0.4$ \\
\includegraphics[align=c,width=1.5in]{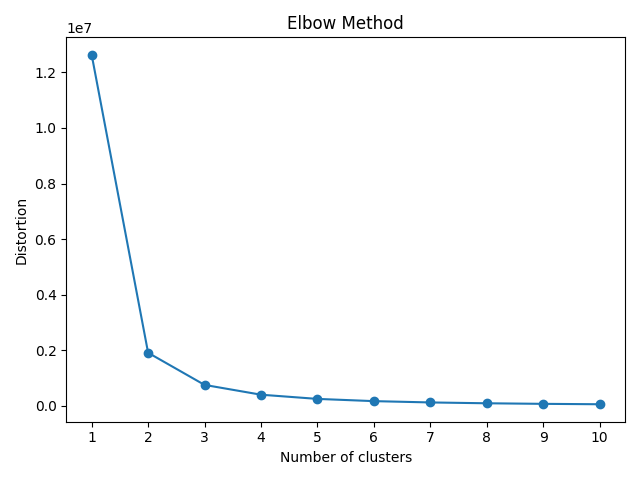}
&
\includegraphics[align=c,width=1.5in]{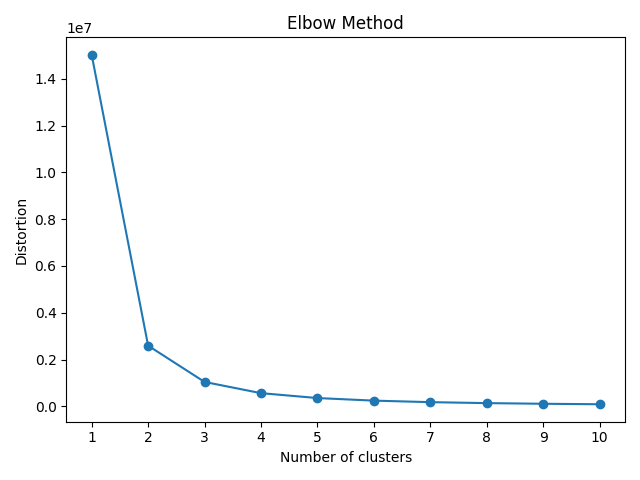}
&
 \includegraphics[align=c,width=1.5in]{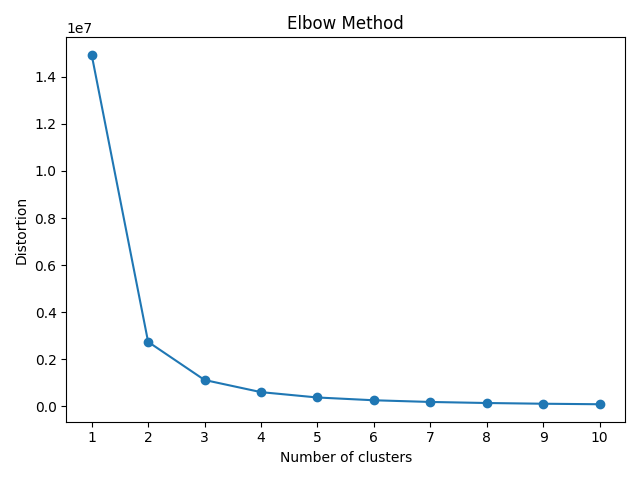}
&
\includegraphics[align=c,width=1.5in]{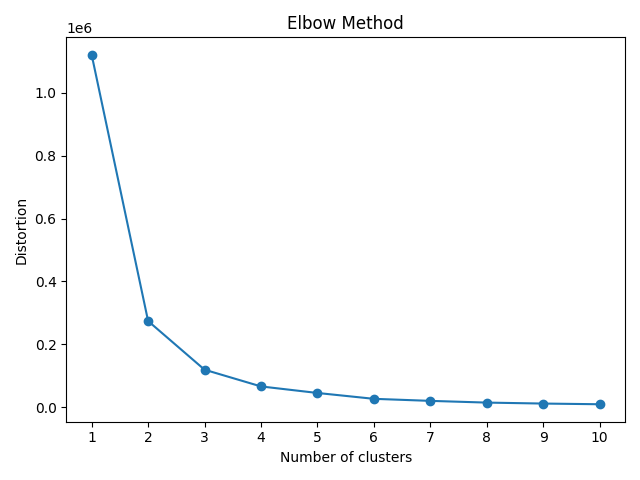}
 
\end{tabular}
\end{center}
\caption{Elbow method plots for information access clustering on the Google Scholar dataset.}
\label{fig:gscholar_elbow}
\end{figure*}

\begin{figure}[htbp]
\begin{center}
\begin{tabular}{c||ccccccccc}
\multicolumn{9}{c}{\textbf{Google Scholar}}\\
& \multicolumn{8}{c}{$k$} \\
$\alpha$ & 2 & 3 & 4 & 5 & 6 & 7 & 8 & 9 & 10\\
\hline
0.01 & \textbf{0.7448} & 0.6919 & 0.6630 & 0.6445 & 0.6317 & 0.6222 & 0.6154 & 0.6088 & 0.6040\\
0.03 & \textbf{0.7154} & 0.6803 & 0.6613 & 0.6489 & 0.6403 & 0.6332 & 0.6264 & 0.6208 & 0.6171\\
0.05 & \textbf{0.7191} & 0.6918 & 0.6782 & 0.6702 & 0.6629 & 0.6590 & 0.6547 & 0.6505 & 0.6478\\
0.40 & \textbf{0.9132} & 0.8993 & 0.8903 & 0.8910 & 0.8851 & 0.8785 & 0.8786 & 0.8782 & 0.8717\\
\end{tabular}
\end{center}
\caption{Silhouette values for varying $\alpha$ and $k$ for the Google Scholar dataset for information access clustering.  The largest silhouette value for each $\alpha$ is shown in bold.}
\label{fig:gscholar_silhouette}
\end{figure}

\end{document}